\documentclass[english,10pt,journal]{IEEEtran}

\usepackage{amssymb,amsmath,amsfonts,amssymb,subfigure,wrapfig}
\usepackage{babel}
\usepackage{setspace,color,algorithm,algorithmic}
\usepackage[unicode=true,pdfusetitle,bookmarks=true,bookmarksnumbered=false,bookmarksopen=false, breaklinks=false,pdfborder={0 0 1},backref=false,colorlinks=false]{hyperref}
\usepackage{breakurl}
\usepackage{multirow}
\usepackage{amsmath,amssymb,mathrsfs,bm}
\usepackage{graphicx}

\begin{document}

\title{Predicting Cyber Attack Rates with Extreme Values}

\author{Zhenxin Zhan, Maochao Xu, and Shouhuai Xu
\thanks{Copyright (c) 2013 IEEE. Personal use of this material is permitted. However, permission to use this material for any other purposes
must be obtained from the IEEE by sending a request to pubs-permissions@ieee.org}
\thanks{Zhenxin Zhan and Shouhuai Xu are with the Department of Computer Science,
University of Texas at San Antonio, San Antonio, TX 78249. Emails: {\tt jankins.ics@gmail.com} (Zhenxin Zhan), {\tt shxu@cs.utsa.edu} (Shouhuai Xu; corresponding author)}
\thanks{Maochao Xu is with the Department of Mathematics, Illinois State University, Normal, IL 61790.
Email: {\tt mxu2@ilstu.edu}}}

\maketitle

\begin{abstract}
It is important to understand to what extent, and in what perspectives, cyber attacks can be predicted.
Despite its evident importance, this problem was not investigated until very recently, when we proposed using
the innovative methodology of {\em gray-box prediction}.
This methodology advocates the use of
gray-box models, which accommodate the statistical properties/phenomena exhibited by the data.
Specifically, we showed that gray-box models that accommodate the Long-Range Dependence (LRD) phenomenon
can predict the attack rate (i.e., the number of attacks per unit time) 1-hour ahead-of-time with an accuracy of 70.2-82.1\%.
To the best of our knowledge, this is the first result showing the feasibility of prediction in this domain.
We observe that the prediction errors are partly caused by the models' incapability in predicting the large attack rates,
which are called {\em extreme values} in statistics.
This motivates us to analyze the {\em extreme-value phenomenon},
by using two complementary approaches: the Extreme Value Theory (EVT) and the Time Series Theory (TST).
In this paper, we show that EVT can offer long-term predictions (e.g., 24-hour ahead-of-time),
while gray-box TST models can predict attack rates 1-hour ahead-of-time with an accuracy of 86.0-87.9\%.
We explore connections between the two approaches, and point out future research directions.
Although our prediction study is based on specific cyber attack data, our methodology can be equally applied to analyze
any cyber attack data of its kind.
\end{abstract}

\begin{IEEEkeywords}
Extreme values, extreme value theory, prediction, gray-box models, time series
\end{IEEEkeywords}

\IEEEpeerreviewmaketitle

\section{Introduction}

Data-driven cybersecurity analytics can deepen our understanding about the statistical phenomena/properties of cyber attacks,
and can potentially help predict cyber attacks (at least from certain perspectives).
Being able to predict cyber attacks, even for minutes (if not hours) ahead-of-time, would
allow the defender to proactively allocate resources for adequate defense
(e.g., dynamically allocating sufficient resources for deep packet inspection or flow-level assembly and analysis).
Despite its apparent importance, there has been little progress in the prediction of cyber attacks until very recently in \cite{XuIEEETIFS13},
perhaps because of the lack of real data and the lack of readily usable prediction models.

In \cite{XuIEEETIFS13}, we proposed using the methodology of {\em gray-box prediction} to predict the cyber attack rate,
namely the number of attacks per unit time.
Gray-box prediction uses {\em gray-box models} that can accommodate the statistical properties/phenomena exhibited by the data.
In contrast, {\em black-box models} do not care about the statistical properties/phenomena exhibited by the data.
We found for the first time that the Long-Range Dependence (LRD) property/phenomenon is exhibited by
honeypot-collected cyber attack data \cite{XuIEEETIFS13}.
For three real data sets that will be called Periods I, II and III, the gray-box models accommodating the LRD phenomenon
can predict attack rates 1-hour ahead-of-time with an accuracy of 82.1\% (error: 17.9\%), 78.3\% (error: 21.7\%) and 70.2\% (error: 29.8\%), respectively.
In contrast, the black-box models, which cannot accommodate the LRD phenomenon,
can only predict attack rates 1-hour ahead-of-time with an accuracy of 55.4\%, 63.7\% and 72.7\%, respectively.
This means that gray-box models can predict more accurately than black-box models.
While this result is encouraging, we need to further improve the prediction accuracy of gray-box models to make them practically employable.
For this purpose, we observe that the aforementioned 17.9\%, 21.7\% and 29.8\% prediction errors of gray-box models
can be partly attributed to their inability to predict the large attack rates,
which are called {\em extreme values} in statistics.
Since extreme values are pervasive in the data, we analyze the
{\em extreme-value} phenomenon in hope of achieving substantially more accurate predictions.

\subsection{Our Contributions}

We propose a methodology for predicting attack rates in the presence of extreme values.
The methodology analyzes the extreme-value phenomenon via two complementary approaches:
the Time Series Theory (TST) and the Extreme Value Theory (EVT).
Although our case study is based on specific cyber attack data collected by a honeypot,
the methodology can be equally applied to analyze any cyber attack data of its kind
(e.g., attacks against production networks).
Specifically, we make two contributions.

First, for short-term predictions, we propose a family of gray-box TST models, called FARIMA+GARCH models,
to accommodate the LRD and extreme-value phenomena exhibited by the data.
For the three real data sets mentioned above,
these gray-box models can predict attack rates 1-hour ahead-of-time at an accuracy of 86.2\% (error: 13.8\%), 87.9\% (error: 12.1\%) and 86.0\% (error: 14.0\%), respectively.
Compared with the gray-box FARIMA models that can accommodate the LRD phenomenon but not the extreme-value phenomenon \cite{XuIEEETIFS13},
the accommodation of extreme values allows FARIMA+GARCH to predict attack rates at an {\em improved} accuracy of 4.1\%, 9.6\% and 15.8\%, respectively.
We also show that the gray-box FARIMA+GARCH models offer substantially more accurate predictions than Hidden Markov Models (HMM) and Symbolic Dynamics (SD) models.
Despite that there are two other data sets for which FARIMA+GARCH cannot predict as accurately as we want,
FARIMA+GARCH predictions are still substantially more accurate than the other models.
Therefore, this model's inadequacy does not invalidate the gray-box prediction methodology;
rather, it suggests further enhancing these FARIMA+GARCH models to accommodate the other statistical properties/phenomena,
which have yet to be identified but are exhibited by those two data sets.

Second, we show that the EVT and TST approaches should be used together in practice because
EVT-based methods are more appropriate for long-term predictions (e.g., 24-hour ahead-of-time)
and TST-based methods are more appropriate for short-term predictions (e.g., 1-hour ahead-of-time).
A combination of the two approaches can lead to more accurate results.
On one hand, the defender can
allocate defense resources utilizing EVT-based predictions of the magnitudes of attack rates (which are predicted 24-hour ahead-of-time),
while making adjustments incorporating TST-based predictions of maximum attack rates (which are predicted 1-hour ahead-of-time).
With a certain degree of agility, the defender can allocate defense resources
with respect to TST-based predictions of the maximum attack rates,
while taking into account EVT-based predictions of the magnitudes of attack rates.

\subsection{Related Work}

To the best of our knowledge, the problem of predicting cyber attacks has not be investigated in the literature.
This is perhaps because of (i) the rule of thumb that
cyber attacks are not predictable, (ii) the lack of real data, and/or (iii) the lack of readily usable prediction methods.
However, we recently showed that cyber attack rates can be predicted
1-hour ahead-of-time via gray-box models, which accommodate the LRD phenomenon that is exhibited by the data  \cite{XuIEEETIFS13}.
Noticing that the prediction errors in \cite{XuIEEETIFS13} are partly caused by the model's failure in predicting large attack rates,
we analyze the extreme-value phenomenon in hope to substantially improve the prediction accuracy.
To the best of our knowledge, this is the first study on analyzing the extreme-value phenomenon in cyber attack data and the first study on predicting cyber attack rates
in the presence of extreme values.

While we believe that studying how to predict cyber attack rates is important on its own, it is also important to make  predictions practical.
This issue is sometimes called the {\em semantic gap} \cite{Sommer:2010:OCW:1849417.1849980}.
We note that the practical use of gray-box prediction models is dependent upon the nature of the data,
and that the data we use for our case study in the present paper is collected by a honeypot.
In order to maximize the practical use of predictions, one might consider
blending honeypot IP addresses into production network IP addresses and frequently randomizing them.
This would make it hard for attackers to identify those honeypot IP addresses and then evade from them.
It is worth mentioning that both extreme values and anomaly behaviors are statistical outliers, but of different kinds.
That is, their treatment requires different methodologies.

Since our investigation is based on honeypot-collected data,
we should mentioned that there is a rich body of literature on analyzing honeypot data from perspectives such as:
(i) distinguishing known vs. unknown attacks \cite{new.att.honeypot},
(ii) identifying traffic characteristics \cite{att.framework,honeypot.forensics,iat.cliques,vis.att,honeypot.pca,cluster.cliques},
(iii) extracting probing and scan activities \cite{PaxsonIEEETIFS2011,flow.nids},
(iv) characterizing Denial-of-Service (DoS) attacks \cite{flow.dos},
(v) studying worm and botnet activities \cite{flow.worm,flow.worm.05,flow.botnet,flow.botnet.WoNS},
and others \cite{honeypot.router,eurecom.attev,nonsta.norm}.
However, prediction is the main task of the present paper.

Finally, there are investigations that are loosely related to ours, such as
the analysis of blackhole-collected data (e.g., \cite{PangIMC2004,BaileyIMC2010}) and one-way traffic \cite{FontasIMC2012}.
These studies aim to classify the data into  classes (e.g. scanning, peer-to-peer applications, unreachable services, misconfigurations, worms etc).
A more recent study of blackhole-collected data aims to characterize the cybersecurity posture  \cite{XuInTrust2014}.
However, these studies do not consider the issue of prediction, which is our perspective.

The paper is organized as follows. Section \ref{sec:preliminaries} briefly reviews
some preliminary statistical knowledge.
Section \ref{sec:methodology} describes our statistical analysis methodology.
Section \ref{sec:evt-analysis} uses EVT to analyze extreme attack rates.
Section \ref{sec:time-series} uses TST to analyze the time series data.
Section \ref{sec:connection} compares the prediction power of multiple methods, and
explores connections between EVT-based and TST-based predictions.
Section \ref{sec:limitations} discusses limitations of the present study and directions for future research.
Section \ref{sec:conclusion} concludes the paper.

\section{Statistical Preliminaries}
\label{sec:preliminaries}

We now review the statistical concepts and techniques that are used in the present paper.
Throughout the paper, we use the more intuitive terms ``outliers'', ``extreme values'', ``extreme events'', ``extreme-value events'' and
the EVT jargon ``exceedances'' interchangeably.
We also use the intuitive term ``average inter-arrival time" between consecutive extreme values and the EVT jargon ``return period" interchangeably.

\subsection{Statistics of the Extreme-Value Phenomenon}
\label{ex-index}

\begin{figure}[!hbtp]
\centering
\includegraphics[width=.48\textwidth]{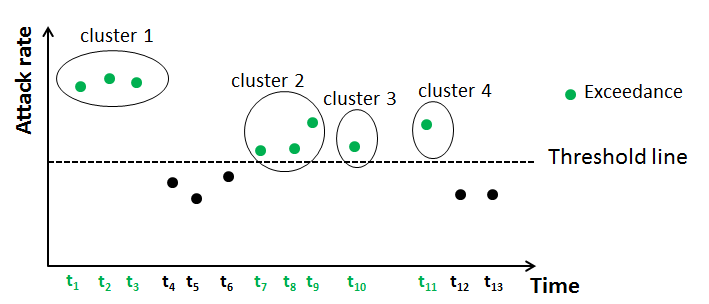}
\caption{Illustration of the extreme-value phenomenon: the dashed line represents a threshold;
green dots are extreme values or {\em exceedances};
black dots are non-extreme values;
extremal index $\theta$ indicates the clustering degree of extreme values (where a cluster is a set of consecutive extreme values).
\label{fig:extremal-index-show-figure}}
\end{figure}

Figure \ref{fig:extremal-index-show-figure} illustrates a time series of attack rates (per  unit time),
where the threshold line corresponds to a threshold value $\mu$ such that the green dots (above the threshold line) are {\em extreme attack rates} or {\em extreme values}.
At a high level, the extreme-value phenomenon can be characterized from a ``spatial'' perspective (i.e., the {\em distribution} of the magnitude of extreme values),
a ``temporal'' perspective (i.e., the {\em inter-arrival time} between two consecutive extreme values), and a ``spatial-temporal''
perspective (i.e., the concept of {\em return level} described below).

\subsubsection{Distribution of extreme values}
\label{gpd:def}

It is known that if $X_1,\ldots,X_n$ are stationary,
then $[X_i-\mu|X_i>\mu]$ may follow the standard Generalized Pareto Distribution (GPD) with survival function
$$\bar G_{\xi,\sigma(\mu)}(x)=1-  G_{\xi,\sigma(\mu)}=\left\{\begin{array}{cc}
  \left(1+\xi \dfrac{x}{\sigma}\right)^{-1/\xi}, & \xi\ne 0, \\
  \exp\{-x/\sigma\}, & \xi=0.
\end{array}\right.,
$$
where $x\in \mathbb{R}^{+}$ if $\xi\in\mathbb{R}^+$ and $x\in [0,-\sigma/\xi]$ if $\xi\in \mathbb{R}^{-}$,
and $\xi$ and $\sigma$ are respectively called the {\em shape} and {\em scale} parameters.
If $X_1,\ldots,X_n$ are from a non-stationary process, then $[X_i-\mu|X_i>\mu]$
may follow a non-stationary GPD with time-dependent parameters, namely
$$\bar G_{\xi(t),\sigma(t)}(x)=\left\{\begin{array}{cc}
  \left(1+\dfrac{\xi(t) x}{\sigma(t)}\right)^{-1/\xi(t)}, & \xi(t)\ne 0, \\
  \exp\{-x/\sigma(t)\}, & \xi(t)=0.
\end{array}\right.
$$
In order to know which of these two scenarios we encounter,
we use the Point Over Threshold (POT) method to fit extreme values via these distributions \cite{Em1997,Re07}.

\subsubsection{Extremal index $0< \theta\leq 1$}
This index reflects the degree at which the extreme values are clustered (cf. Figure \ref{fig:extremal-index-show-figure}):
$1/\theta$ indicates the mean size of clusters (i.e., the mean number of extreme values in a cluster),
where $\theta=1$ implies that each cluster has one extreme value (i.e., there is no clustering phenomenon).
Formally, let
$\{X_1, X_2,\ldots,\}$ be a sequence of random variables from a stationary process.
Let $M_n=\max \{X_1,\ldots, X_n\}$. Under certain regularity conditions, it holds that
$$\lim_{n\rightarrow\infty} P\left(\frac{M_n-b_n}{a_n}\le x\right)=H_\xi^\theta(x),$$
where $a_n$ and $b_n$ are normalizing constants,
\begin{equation}\label{extreme}
H_\xi(x)=\left\{\begin{array}{cc}
         \exp\left\{-(1+\xi \frac{x-\mu}{\sigma})^{-1/\xi}\right\}, & \xi\ne 0 \\
          \exp\{-e^{-\frac{x-\mu}{\sigma}}\},& \xi=0
       \end{array}\right.
\end{equation}
is a non-degenerate distribution function with $1+\xi \frac{x-\mu}{\sigma}>0$, and $\theta\in (0,1]$ is the {\em extremal index}.
We will characterize the extremal index when fitting the extreme attack rates.

\subsubsection{Return level}
This index reflects the expected {\em magnitude} of extreme values (but not necessarily the {\em maximum} value).
Let $T$ be the average inter-arrival time between two consecutive extreme values,
which is also called the {\em return period}.
The probability that an extreme event occurs is $p=1/T$.
The concept of {\em return level} identifies a special threshold value such that there is, on average, a single extreme event during each return period.
Formally, suppose random variable $X$ has a stationary GPD with shape parameter $\xi$ and scale parameter $\sigma$.
Then,
$$P(X>x)=\zeta_\mu \left[1+\xi\left(\frac{x-\mu}{\sigma}\right)\right]^{-1/\xi}, ~~~\xi\ne 0,$$
where $\zeta_\mu=P(X>\mu)$. The {\em return level} $x_m$ is exceeded (on average) once per $m$ observations (i.e., time intervals), and is given by
$x_m=\mu+\sigma/\xi \left[(m\zeta_\mu)^\xi-1\right]$.
For non-stationary GPD, the return level is given by  $x_m=\mu+\sigma(m)/\xi(m) \left[(m\zeta_\mu)^{\xi(m)}-1\right]$.
We will use this method to predict the return level of extreme attack rates.

\subsection{Properties and Models of Time Series}
\label{sec:tst}

\subsubsection{Long-Range Dependence (LRD)}
Unlike the Poisson process with the memoryless property,
a LRD process exhibits that the autocorrelation decays slower than the exponential decay.
Formally, a stationary time series $\{X_t\}$ exhibits LRD if its autocorrelation
$r(h)={\rm Cor}(X_i,X_{i+h})\sim h^{-\beta} L(h)$ as $h\rightarrow\infty$,
where $0<\beta<1$, and $L(\cdot)$ is a slowly varying function with $\lim\limits_{x\rightarrow \infty} \frac{L(tx)}{L(x)}=1$ for all $t>0$ \cite{Em1997}.
Note that $\sum_{h} r(h)=\infty$.
The degree of LRD is quantified by the Hurst parameter $H=1-\beta/2$,
meaning that $1/2<H<1$ and the degree of LRD increases as $H\to 1$.
LRD is exhibited by the data, which is analyzed in \cite{XuIEEETIFS13}
and further analyzed in the present paper.

\subsubsection{FARIMA and GARCH Time Series Models}
\label{sec:ts}

FARIMA (Fractional AutoRegressive Integrated Moving Average)
and GARCH (Generalized AutoRegressive Conditional Heteroskedasticity) are two widely used time series models \cite{CC2008}.
FARIMA can accommodate LRD,
while GARCH can accommodate the extreme-value phenomenon.
We will use FARIMA+GARCH models to predict attack rates.
Specifically, let $\phi(x)=1-\sum_{j=1}^p \phi_j x^j$, $\psi(x)=1+\sum_{j=1}^q \psi_j x^j$,
and $\epsilon_t$ be independent and identical normal random variables with mean $0$ and variance $\sigma^2_\epsilon$.
A time series $\{X_t\}$ is called a FARIMA$(p,d,q)$ process if
$\phi(B)(1-B)^d X_t=\psi(B) \epsilon_t$,
where $-1/2<d<1/2$, and $B$ is the back shift operator with $B X_t=X_{t-1}$, $B^2 X_t=X_{t-2}$, etc.
A time series $\{X_t\}$ is called a GARCH process \cite{bll2010} if
$X_t=\sigma_t \epsilon_t$, where $\epsilon_t$ (also called {\em innovation}) is the standard white noise.
We consider two variants of GARCH. For the Standard GARCH (SGARCH) model, we have
$\sigma_t^2=w+\sum_{j=1}^q \alpha_j \epsilon_{t-j}^2+\sum_{j=1}^p \beta_j\sigma^2_{t-j}$.
For the Integrated GARCH (IGARCH) model, we have
$\phi(B)(1-B)\epsilon_t^2=w+(1-\psi(B))v_t$,
where $v_t=\epsilon_t^2-\sigma_t^2$.
To accommodate more general classes of noise, we will use the
skewed Student-T distribution (SSTD) or skewed Generalized Error distribution (SGED).

\subsubsection{Hidden Markov Model (HMM) for Time Series}
\label{sec:hmm}
HMM can describe a Markov process with hidden states that are not directly observable.
Specifically, suppose $x_t$ is the observation of some Markov process with hidden state $s_t$ at time $t$.
The joint distribution of a sequence of states and observations can be written as
   $$p(x_{1},\ldots,x_{T})=p(s_1)p(x_1|s_1)\prod_{t=2}^T p(s_t|s_{t-1}) p(x_t|s_t),$$
where $p(s_1)$ is the probability of the initial state $s_1$, and $p(s_t|s_{t-1})$ is the state transition probability.
It is often assumed that $p(x_t|s_t)$ follows the Gaussian distribution \cite{zu2009hidden}.

\subsubsection{Symbolic Dynamics (SD) Model of Time Series}
\label{sec:sd}
The basic idea underlying symbolic time series analysis is the following: partitioning the phase space,
encoding the observed time series data into nonlinear system dynamics, and constructing a finite state
machine model from the symbol sequence.
For example, a time series  $\{X_t: t=1,\ldots, T\}$ may be reduced to a string $\{\bar X_1,\ldots, \bar X_w\}$,
where $w< T$, and $\bar X_i$ is a symbol representing some aggregation as an approximation to the original time series \cite{lin2003symbolic}.

\section{Data and Analysis Methodology}
\label{sec:methodology}

\subsection{Honeypot}

A honeypot is a monitored instrument for passively collecting probe and attack traffic \cite{provos:2004}.
A honeynet is composed of multiple honeypots \cite{honeynet}.
In general, there are two types of honeypots.
High-interaction honeypots are systems with real vulnerabilities in their service programs.
This instrument is not scalable because it consumes a lot of computing resources.
Moreover, running high-interaction honeypots often involves legal issues, which is true at least in the United States.
Low-interaction honeypots usually simulate software vulnerabilities.
This instrument is scalable because it conducts limited interactions with attackers.
The data we analyze in the present data is collected by a low-interaction honeypot, which consists of 166 consecutive IP addresses.
The honeypot runs four honeypot programs:
Amun \cite{amun}, Dionaea \cite{dionaea}, Mwcollector \cite{mwcollect} and Nepenthes \cite{nepenthes}.
Each program is associated to a unique IP address. Each physical computer monitors multiple IP addresses.

\subsection{Data}
\label{sec:data-description}

The data we analyze is the same as in \cite{XuIEEETIFS13}.
This is plausible because our predictions of attack rates directly improve upon those in \cite{XuIEEETIFS13}.
The data was collected during five periods of time in 2010-2011, and these five periods of time correspond to 47, 18, 54, 21 and 80 days, respectively.
We extract the attack traffic from raw data (in {\tt pcap} format) with respect to the ports that are monitored by those honeypot programs.
We treat each TCP flow initiated by a remote computer as an attack because the honeypot does not offer any legitimate service.
An unsuccessful TCP handshake is also deemed as attack because a handshake could have been dropped by a honeypot program.
For assembling TCP flows, we set the flow lifetime as 300 seconds (i.e., an attack/flow does not span over 300 seconds)
and the flow timeout time as 60 seconds (i.e., an attack/flow expires if there is no activity for 60 seconds);
these parameters have been widely used \cite{honeypot-PCA}.

\begin{figure}[!hbtp]
\centering
\centering
\subfigure[Period I]{\includegraphics[width=.24\textwidth]{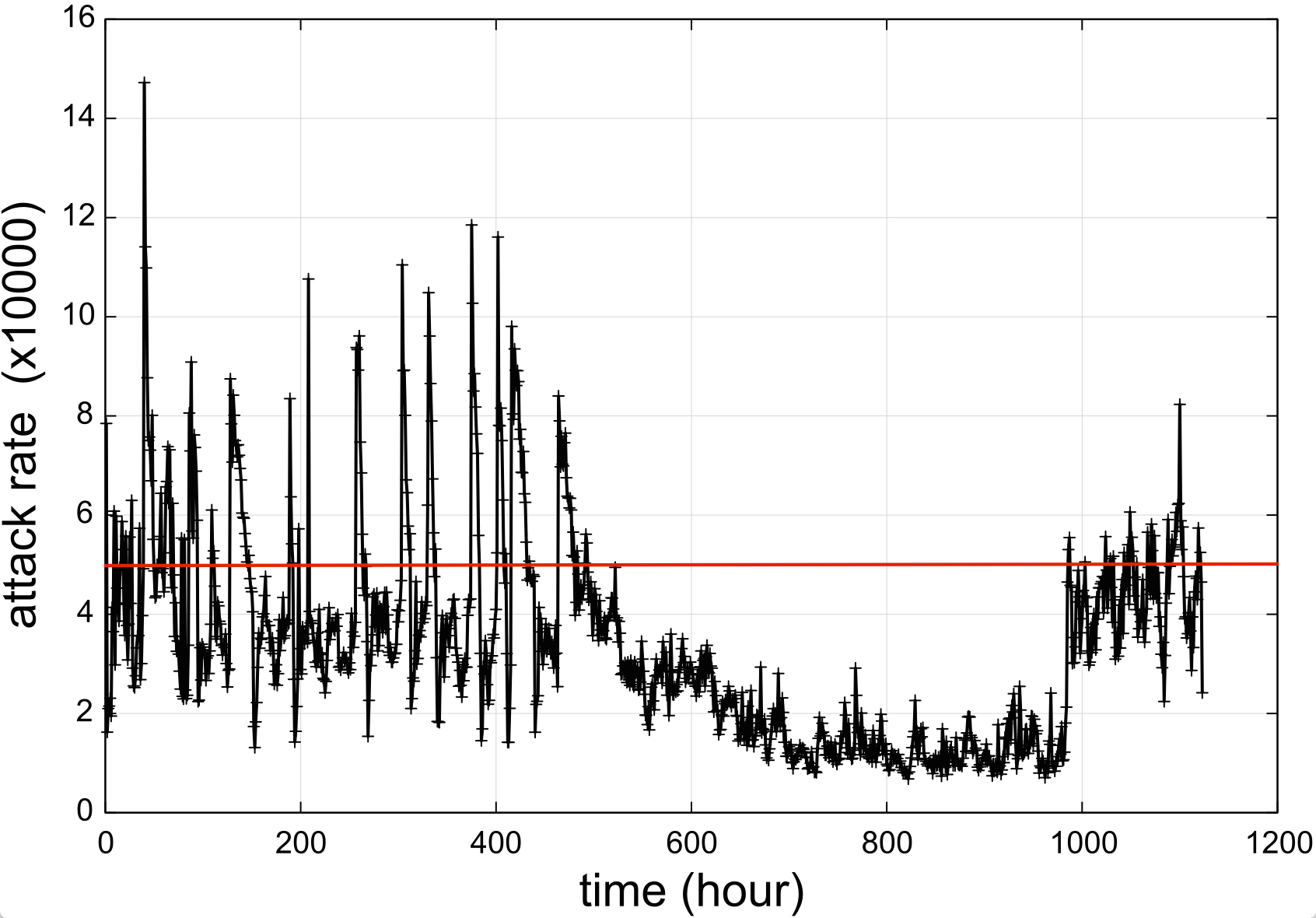}}
\subfigure[Period II]{\includegraphics[width=.24\textwidth]{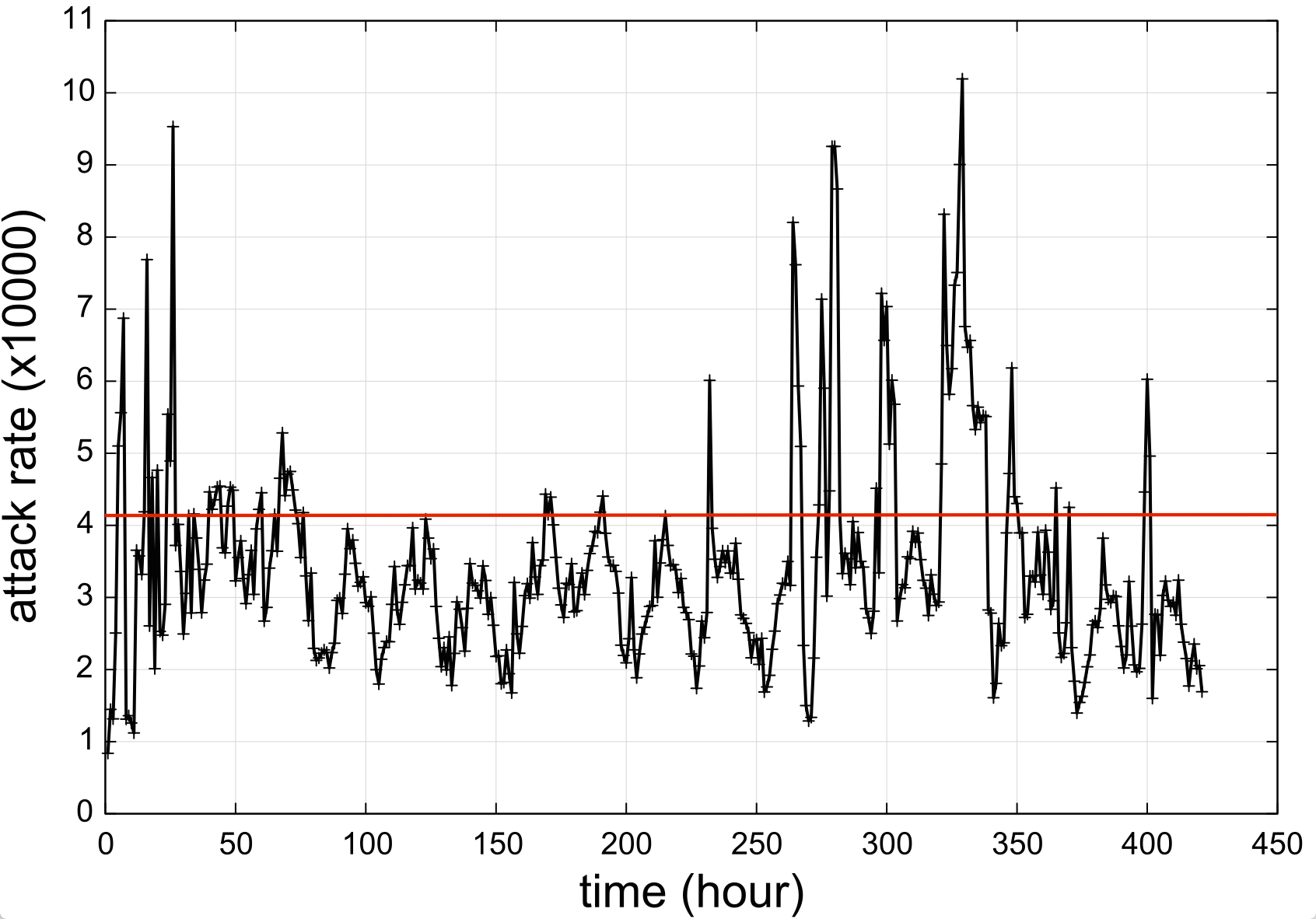}}
\subfigure[Period III]{\includegraphics[width=.24\textwidth]{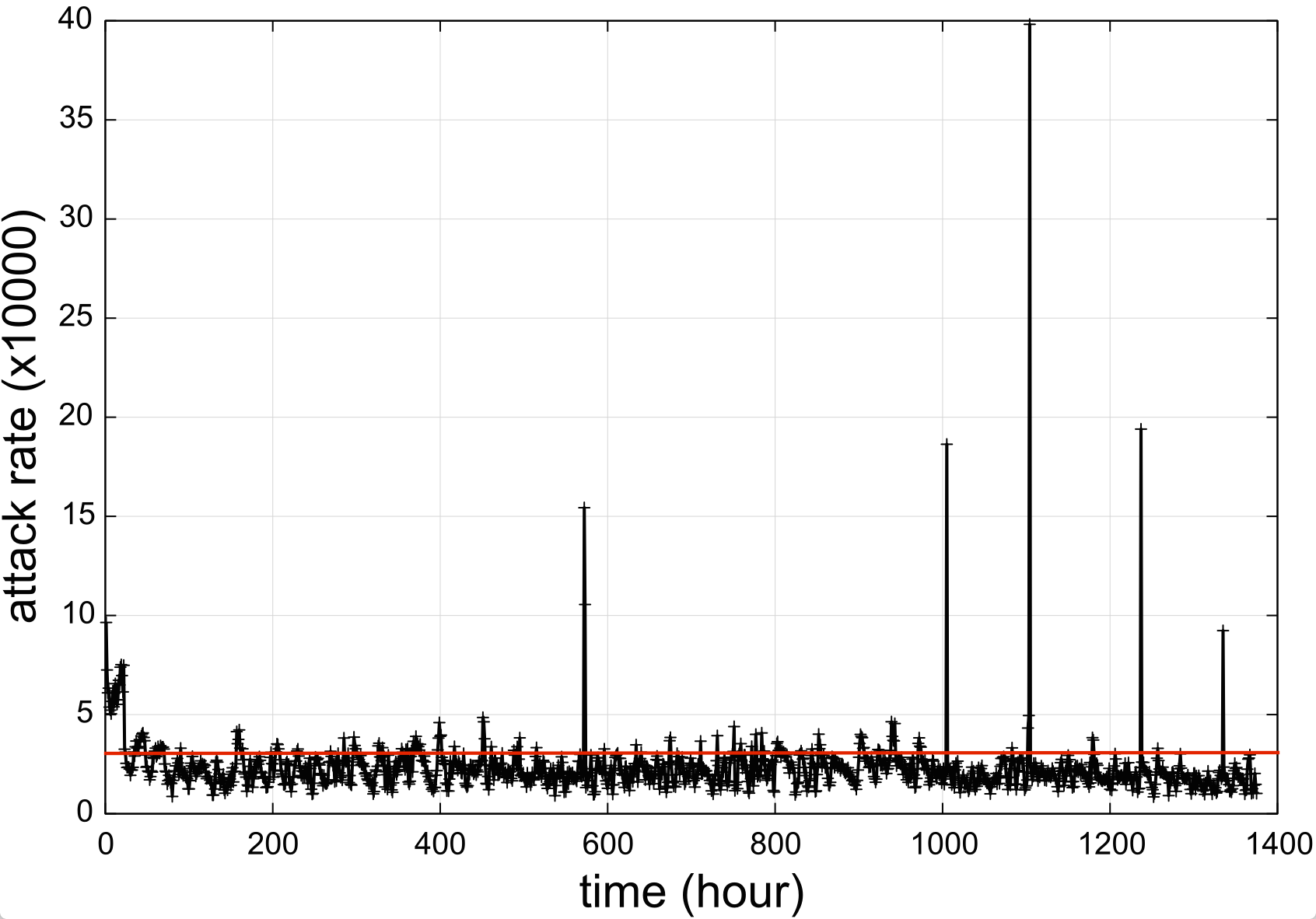}}
\subfigure[Period IV]{\includegraphics[width=.24\textwidth]{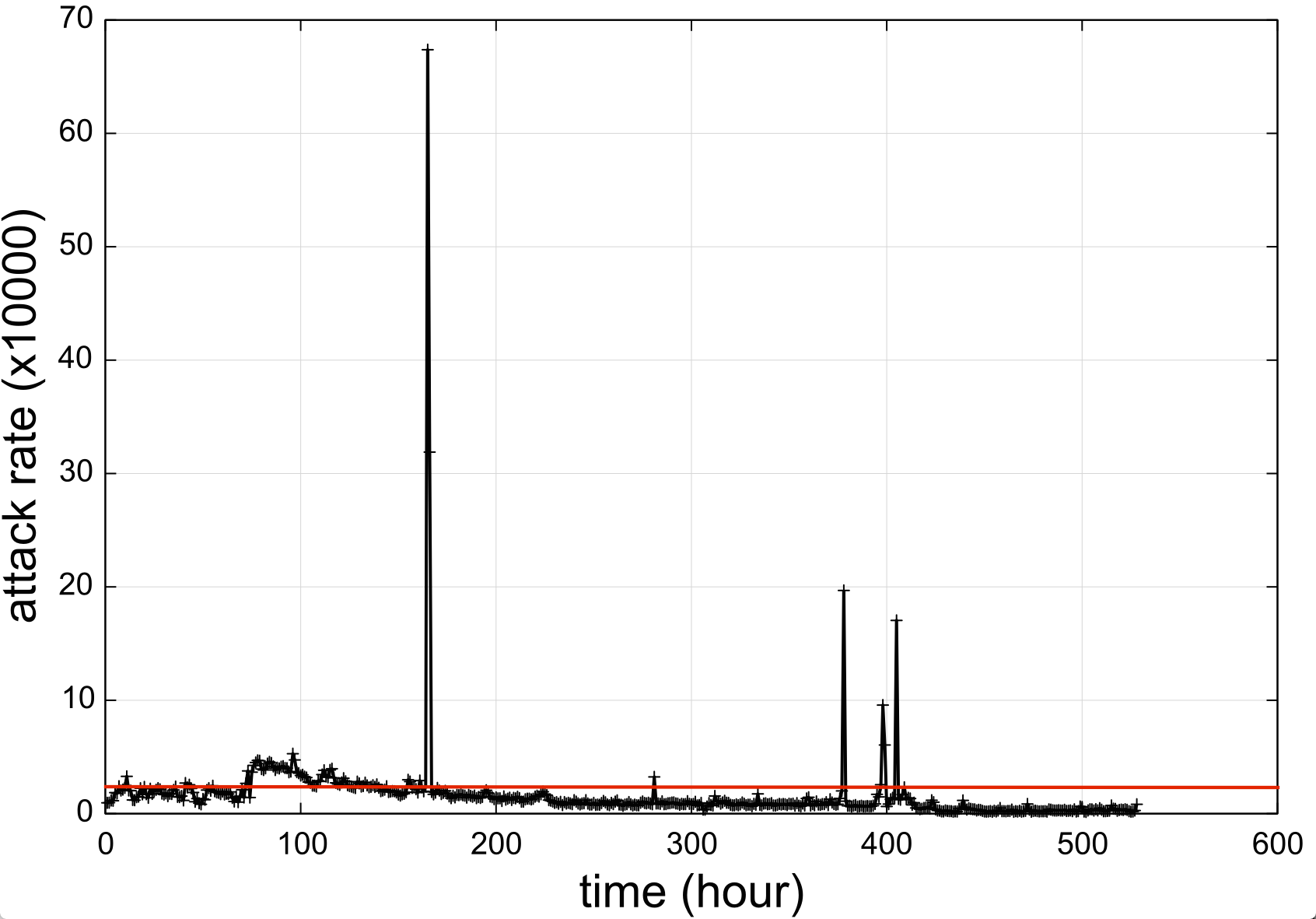}}
\subfigure[Period V]{\includegraphics[width=.24\textwidth]{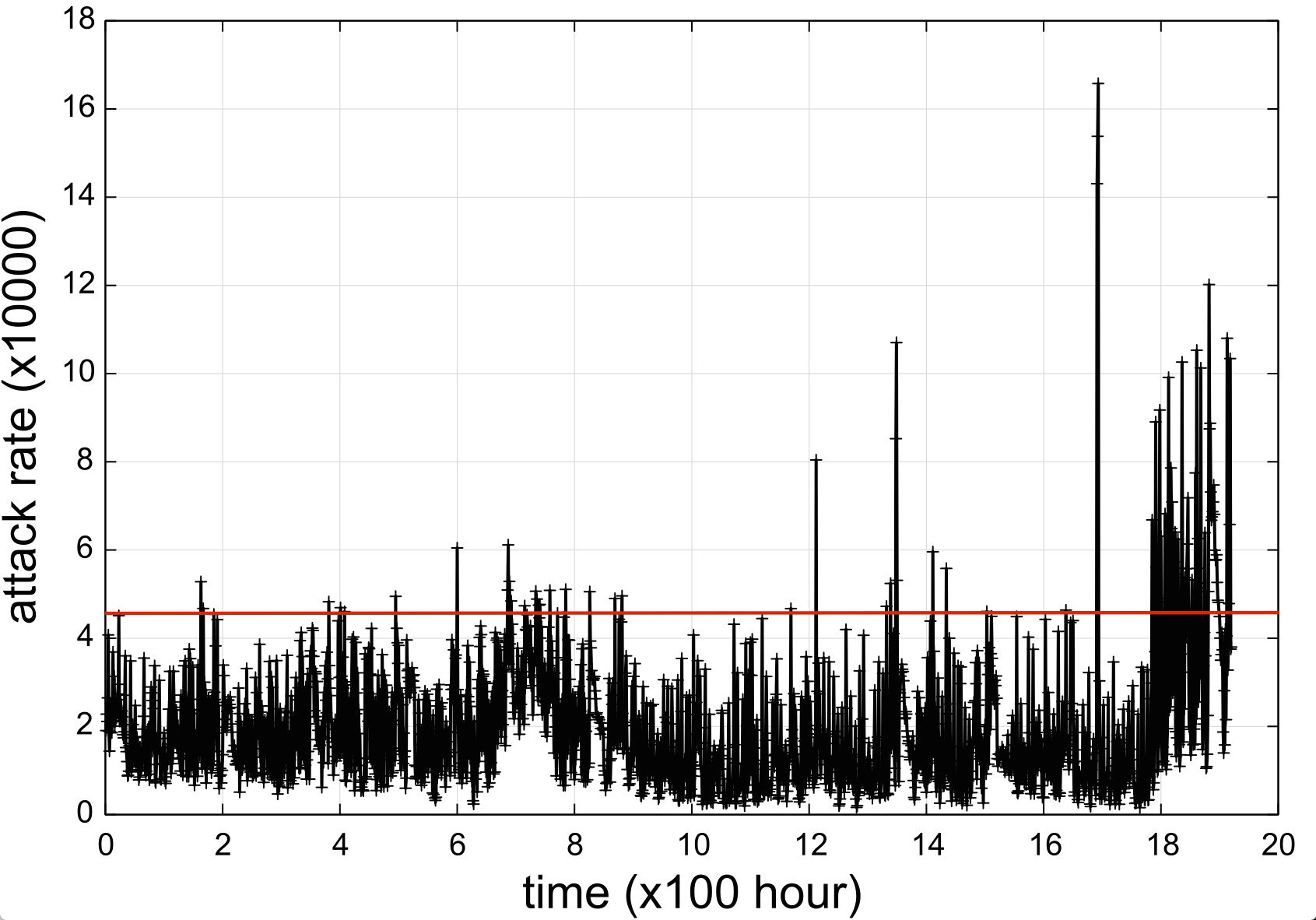}}
\caption{Time series plots of attack rates (number of attacks per hour).
The red lines are some thresholds, and the attack rates above these thresholds are extreme values.
\label{fig:D1-D2}}
\end{figure}

\subsection{The Extreme-Value Phenomenon}

Figure \ref{fig:D1-D2} plots the attack data during the five periods of time, with some illustrative thresholds above which attack rates are considered extreme values.
We observe the extreme-value phenomenon, namely the many extreme attack rates (i.e., spikes) caused by intense attacks.
For example, the two spikes at the 1,237th and 1,335th hours in Period III are SSH traffic (or ``SSH scans");
the spike at the 1,693th hour in Period V is SIP INVITE traffic (or DDoS attacks via SIP INVITE messages).

\subsection{An Extreme-Value Analysis Methodology}
\label{study-methodology}

Our methodology is centered on analyzing statistical properties of attack rates and exploiting these statistical properties to predict
attack rates. For this purpose, we integrate two complementary statistical approaches.
The first approach is based on EVT (Extreme Value Theory), which deals with extreme attack rates.
This approach is appropriate for relatively long-term prediction of extreme events (e.g., 24-hour ahead-of-time),
because (i) extreme events do not occur often,
and (ii) the analysis only considers extreme attack rates.
The second approach is based on TST (Time Series Theory), which does not differentiate between the extreme values (above the threshold) and the non-extreme values (below the threshold).
These two approaches are complementary to each other because (i) the data/information they use is different (i.e., proper subset vs. superset),
and (ii) the predictions they make are different (i.e., return levels or expected magnitude of extreme attack rates vs. concrete attack rates).
Therefore, it is interesting to seek connections between them.

For both EVT-based and TST-based analyses, we proceed as follows.
First, we identify the statistical properties exhibited by the data, such as
the stationarity of the processes that drive the attacks, and
the clustering behavior exhibited by the extreme values.
Second, we exploit the identified statistical properties to fit the relevant data (i.e., {\em gray-box} fitting).
Although there are generic methods for analyzing the extreme-value phenomenon,
those generic methods are not sufficient for our purpose
because they do not consider the properties exhibited by the cyber attack data we analyze.
This observation motivates us to investigate new statistical techniques that are relevant to the cybersecurity domain
(e.g., a family of FARIMA+GARCH models).
Third, we predict attack rates by using EVT- and TST-based methods (i.e., gray-box prediction),
and explore relationships (especially, the consistency) between their predictions.

\section{EVT-based Extreme-Value Analysis}
\label{sec:evt-analysis}

In order to fit the distribution of extreme values, we need to determine whether they are driven by a stationary process
(i.e., the distribution does not change) or they are driven by a non-stationary process.
This will guide us in how we use time-invariant or time-dependent parameters in the models that we select to fit the data.
For this purpose, we consider four candidate models: $M_1,\ldots,M_4$,
where $M_1$ corresponds to the case of stationary processes and the others correspond to the case of non-stationary processes. Specifically,
\begin{itemize}
  \item $M_{1}$:  The standard GPD (Generalized Pareto Distribution).
  \item $M_{2}$:  GPD with time-invariant shape parameter $\xi$ but time-dependent scale parameter
$\sigma(t)= \exp\left(\beta_0+\beta_1 \log(t)\right)$.
  \item $M_{3}$: GPD with time-invariant scale parameter $\sigma$ but time-dependent shape parameter
$\xi(t)=\gamma_1+\gamma_2 t$.
  \item $M_{4}$: GPD with time-dependent parameters
$\sigma(t)=\exp\left(\beta_0+\beta_1 \log(t)\right)$ and
$\xi(t)=\gamma_1+\gamma_2 t$.
\end{itemize}
Since we do not know the stationarity {\em a priori}, we first use $M_1$ to fit the extreme attack rates.
If $M_1$ cannot fit well, we use non-stationary models $M_2,\ldots,M_4$ to fit the extreme attack rates.
We use some standard goodness-of-fit statistics and QQ-plot for evaluating the quality of fitting.

\subsection{Fitting Stationary Extreme Attack Rates}
\label{sec:stationary}

We use Algorithm \ref{alg:stationary-heavy-tail-id} to fit stationary extreme attack rates.
The algorithm uses QQ-plot and two goodness-of-fit statistics called CM  and AD \cite{CS2001}, where both CM and AD measure  the goodness-of-fit of a distribution.
If the $p$-values of both CM and AD statistics are greater than $.1$ (which is more conservative than the textbook criterion $.05$),
and the QQ-plot also confirms the goodness-of-fit,
then the algorithm concludes that the extreme attack rates are stationary and follow the standard GPD;
otherwise, the extreme attack rates are non-stationary and will be fitted via Algorithm \ref{alg:non-stationary-heavy-tail-id},
which is described below.

{\small
\begin{algorithm}[hbtp!]
\caption{Fitting stationary extreme attack rates via $M_1$}
\label{alg:stationary-heavy-tail-id}
INPUT: attack-rate time series\\
OUTPUT: $M_1$ fitting result
\begin{algorithmic}[1]
\STATE initialize $quantileSet$ \COMMENT{assuring $\geq 30$ extreme values}
\FOR{$q\in quantileSet$ (from the minimum to the maximum in increasing order)}
  \STATE{use the standard GPD to fit the extreme attack rates that are greater than threshold quantile $q$}
  \STATE{evaluate goodness-of-fit statistics CM, AD, QQ-plot}
  \IF{fitting is good}
    \STATE{estimate GPD parameters $(\xi,\sigma)$, extremal index $\theta$}
    \RETURN $(q,\xi,\sigma,\theta)$~~~~\COMMENT{the first successful fitting}
  \ENDIF
\ENDFOR
\RETURN{-1}~~\COMMENT{stationary distribution fitting failed}
\end{algorithmic}
\end{algorithm}
}

A key ingredient in Algorithm \ref{alg:stationary-heavy-tail-id} is the threshold quantile, which
specifies the threshold above which an attack rate is an extreme value.
Specifically, $quantileSet$ is an ordered set of quantiles, where the maximum quantile is
chosen to guarantee that there are at least 30 extreme attack rates (because, as a rule of thumb, 30 is required for the sake of reliable fitting), and
the minimum quantile is $20\%$ different from the maximum quantile with step-length $5\%$.
For example, suppose there are 1000 attack rates (corresponding to observations during 1000 hours).
The maximum threshold quantile is $1-\frac{30}{1000}\times{100\%}=97\%$ and
the minimum threshold quantile is 77\%, leading to $quantileSet=\{77\%, 82\%, 87\%, 92\%, 97\%\}$.
The algorithm starts with threshold quantile 77\%, then 82\% etc. (i.e., in increasing order), and halts on the first successful fitting
(in which case, parameters are obtained) or until after all of the fitting attempts fail (i.e., the process is non-stationary).

Table \ref{table:basic-stat-stationary} summarizes the fitting results using $M_1$.
Extreme attack rates in Periods III and V are from stationary processes, but the other three periods cannot be fitted by $M_1$.
Specifically, Period III has threshold quantile $q=90\%$ (i.e., there are 130 extreme attack rates that are above the 90\% quantile)
and extremal index $\theta=0.60$ (i.e., a cluster contains, on average, $1/.60=1.67$ extreme values,
or extensive attacks sustain for 1.67 hours on average).
Period V has threshold quantile $q=95\%$ and extremal index $\theta=.33$.
Combining the above observations and the fact that the five periods are respectively 47, 18, 54, 21, and 80 days,
we suspect that stationary extreme attack rates may not be observed for a period of time shorter than 50 days,
because Periods III and V correspond to 54 and 80 days, respectively.

\begin{table}[!htbp]
\centering
{\footnotesize
\begin{tabular}{|c|r|r|r|r|r|r|c|}
\hline
Period & $q$ & \# of EV & \# of C & $\theta$ & $\xi$ & $\sigma$  & model \\
\hline
 III & 90\% &  130  &  95  &    0.60 &0.36 & 3778.19 & $M_1$ \\
\hline
 V & 95\%  &  96 & 31 & 0.33 &0.16 & 13553.5  & $M_1$ \\
\hline
 \end{tabular}
\caption{\lowercase{EVT-based fitting of stationary attack rates, where $(q,\xi,\sigma,\theta)$ are output by Algorithm \ref{alg:stationary-heavy-tail-id},
``\# of EV" means ``number of extreme values (i.e., extreme attack rates)",
``\# of C" means ``number of clusters",
parameters $\theta,~\xi,~\sigma$ are described in Section \ref{sec:preliminaries}. The last column indicates the best fitting model.}
\label{table:basic-stat-stationary}}
}
\end{table}

\subsection{Fitting Non-Stationary Extreme Attack Rates}

We use Algorithm \ref{alg:non-stationary-heavy-tail-id} to select the best fitting model for the non-stationary extreme attack rates,
where $quantileSet$ is the same as in Algorithm \ref{alg:stationary-heavy-tail-id}.
We use the AIC (Akaike Information Criterion) statistic \cite{CC2008} and QQ-plot to evaluate the goodness-of-fit,
where AIC reflects the fitting loss (i.e., the smaller the AIC value, the better the fitting).
As a rule of thumb, two AIC values are considered {\em fairly close} when their difference is less than 10.
If a model $M_j\in \{M_2,M_3,M_4\}$ incurs the minimum AIC value that is not fairly close to any of the other two models' AIC values, we choose $M_j$ as the best fitting model;
otherwise, we choose the simpler/simplest model whose AIC value is fairly close to the minimum AIC value
(note that model $M_{2}$ is considered simpler than $M_{3}$, which is simpler than $M_{4}$).

{\small
\begin{algorithm}[hbtp!]
\caption{Fitting non-stationary extreme attack rates}
\label{alg:non-stationary-heavy-tail-id}
INPUT: attack-rate time series whose extreme values cannot be fitted by $M_1$ \\
OUTPUT: fitting result
\begin{algorithmic}[1]
\STATE initialize $quantileSet$ \COMMENT{same as in Algorithm \ref{alg:stationary-heavy-tail-id}}
\FOR{$q\in quantileSet$ (from the minimum to the maximum in increasing order)}
  \STATE{use models  $M_{2}$, $M_{3}$ and $M_{4}$ to fit the attack rates that are greater than threshold quantile $q$}
  \STATE{evaluate goodness-of-fit via AIC and QQ-plot}
  \IF{any of the three models fits well}
  \STATE{choose the model with the minimum AIC value,
         or choose the simpler/simplest model whose AIC value is fairly close to the minimum AIC value}
   \RETURN $(q,{\rm AIC~ value})$ of the selected model $M_j$
  \ENDIF
\ENDFOR
\RETURN -1 \COMMENT{failed in fitting extreme attack rates}
\end{algorithmic}
\end{algorithm}
}

Table \ref{tbl:non-stationary-d1d2} summarizes the fitting results.
The AIC values of these three models for Periods I and II are fairly close,
and thus we choose the simpler model $M_2$.
Since the AIC value of $M_{2}$ in Period IV is the smallest, we choose $M_2$ as the best fitting model for Period IV.
For Period IV, models $M_3$ and $M_4$ have smaller AIC values that are fairly close to each other,
and therefore we choose $M_3$ as the best fitting model.

\begin{table}[!hbtp]
\centering
{\footnotesize
\begin{tabular}{|c|r|r|c|c|c|c|c|}
\hline
Period & $q$ & \# of EV & $M_2$ & $M_3$ & $M_4$ & QQ-plot & model \tabularnewline
\hline
I & 85\% & 168 & 3656 & 3653 & 3656 & $\surd$ & $M_2$\tabularnewline
\hline
II & 80\% & 84 & 1774 & 1774 & 1776 & $\surd$ & $M_2$ \tabularnewline
\hline
IV & 80\% & 105 & 1774 & 2014 & 2016 & $\surd$ & $M_2$\tabularnewline
\hline
\end{tabular}
\caption{EVT-based fitting of non-stationary extreme attack rates, where column $M_j$ ($2\leq j \leq 4$) represents the AIC value of model $M_j$,
``$\surd$'' indicates that QQ-plot confirms fitting well (QQ-plots are omitted for saving space),
and the other notations are the same as in Table \ref{table:basic-stat-stationary}.
}
\label{tbl:non-stationary-d1d2} }
\end{table}

\section{TST-based Extreme-Value Analysis}
\label{sec:time-series}

Now we study how TST-based models can fit the extreme values.
Since the attack-rate time series exhibit the LRD phenomenon \cite{XuIEEETIFS13}
and the extreme-value phenomenon, we need models that can accommodate both.
Since the GARCH model can accommodate the extreme-value phenomenon \cite{Em1997} and
the FARIMA model can accommodate LRD,
we propose to use the following FARIMA+GARCH model:
$$\phi(B)(1-B)^d(y_t-\mu_t)=\psi(B)\epsilon_t,$$
where parameters $\phi$ and $\psi$ are the same as reviewed in Section \ref{sec:ts},
$B$ is the lag operator, $(1-B)^d$ is the LRD process with Hurst parameter $H$ satisfying $0<d=H-.5<1$,
and $\mu_t=\mu+\xi \sigma_t$ such that the variance $\sigma_t$ follows either the SGARCH (i.e., Standard GARCH)
or the IGARCH (i.e., Integrated GARCH) with noise distribution SSTD or SGED (as reviewed in Section \ref{sec:ts}).
This actually leads to a family of FARIMA+GARCH models:
\begin{itemize}
\item $M'_1$: FARIMA+SGARCH+SSTD;
\item $M'_2$: FARIMA+SGARCH+SGED;
\item $M'_3$: FARIMA+IGARCH+SSTD;
\item $M'_4$: FARIMA+IGARCH+SGED.
\end{itemize}

For comparison, we also consider the gray-box FARIMA model, which can accommodate the LRD phenomenon but not the extreme-value phenomenon.
Recall that FARIMA can predict the time series more accurately than the LRD-less (i.e. {\em black-box}) ARMA model \cite{XuIEEETIFS13}.
To select the best fitting model, we use two model selection criteria: PMAD (Percent Mean Absolute Deviation) and AIC.
To select the best prediction model, we use PMAD.
Suppose $X_m,\ldots,X_h$ are the observed attack rates and $X'_m,\ldots,X'_h$ are the fitted (predicted) attack rates.
We have ${\rm PMAD}=\sum_{t=m}^{m+h} |X_{t}-X'_{t}|/\sum_{t=m}^{m+h} X_t$,
which captures the overall fitting (prediction) error. Note that the smaller the PMAD value, the better the fitting (prediction).

\begin{figure*}[!hbtp]
\centering
\subfigure[Period I]{\includegraphics[width=.192\textwidth]{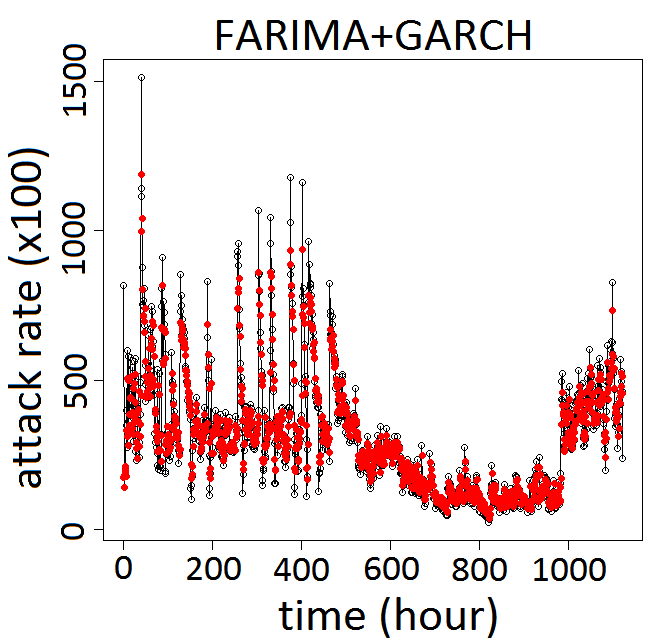}}
\subfigure[Period II]{\includegraphics[width=.192\textwidth]{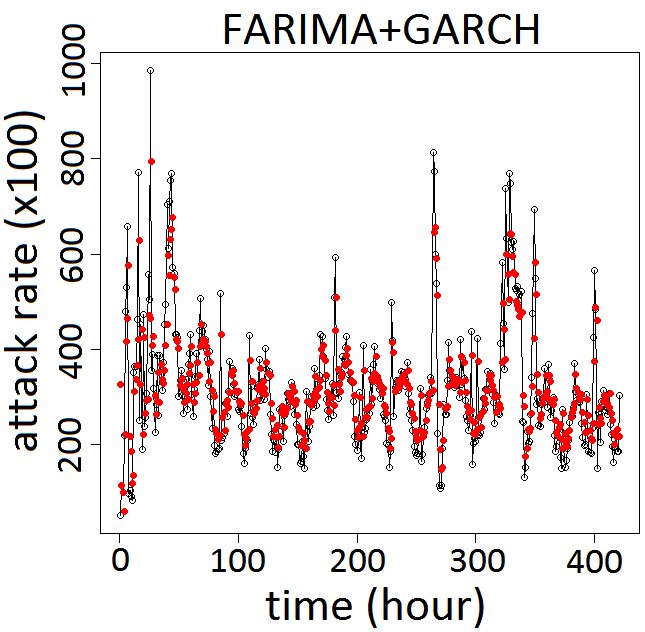}}
\subfigure[Period III]{\includegraphics[width=.192\textwidth]{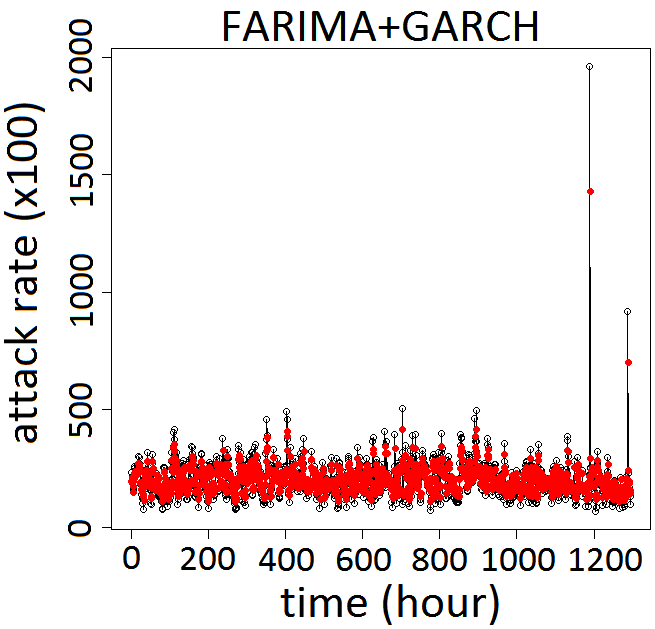}}
\subfigure[Period IV]{\includegraphics[width=.192\textwidth]{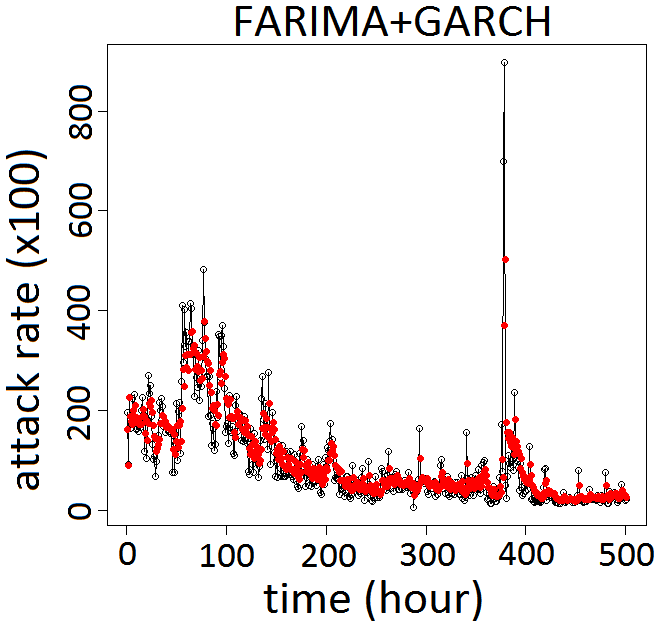}\label{fig:fitting-d2-5d}}
\subfigure[Period V]{\includegraphics[width=.192\textwidth]{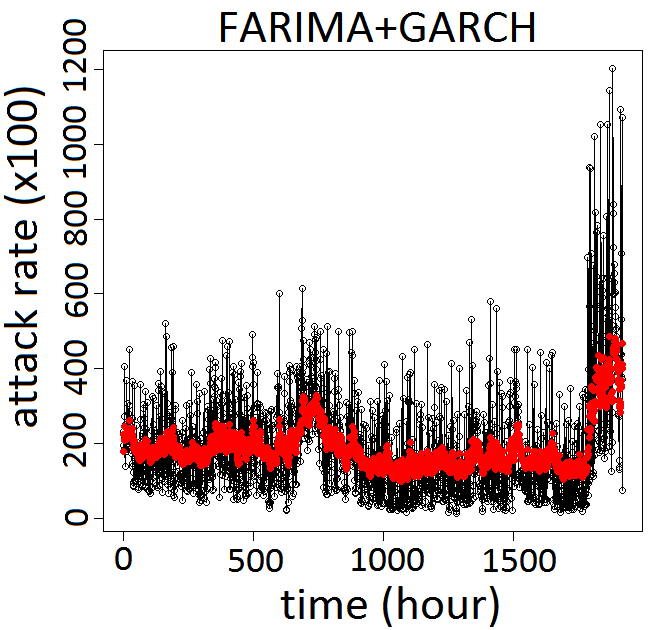}}
\subfigure[Period I]{\includegraphics[width=.192\textwidth]{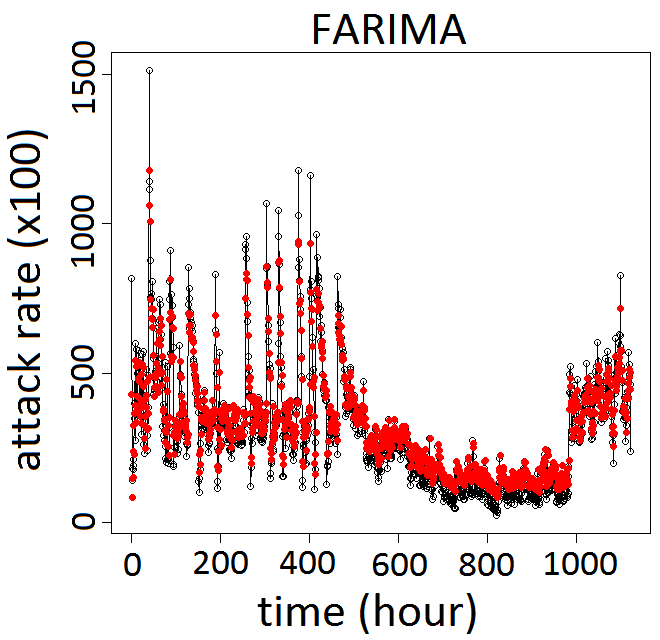}}
\subfigure[Period II]{\includegraphics[width=.192\textwidth]{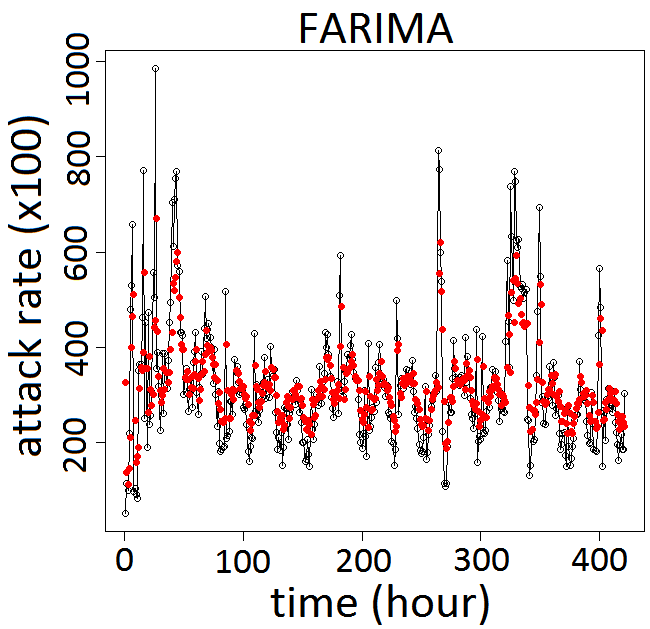}}
\subfigure[Period III]{\includegraphics[width=.192\textwidth]{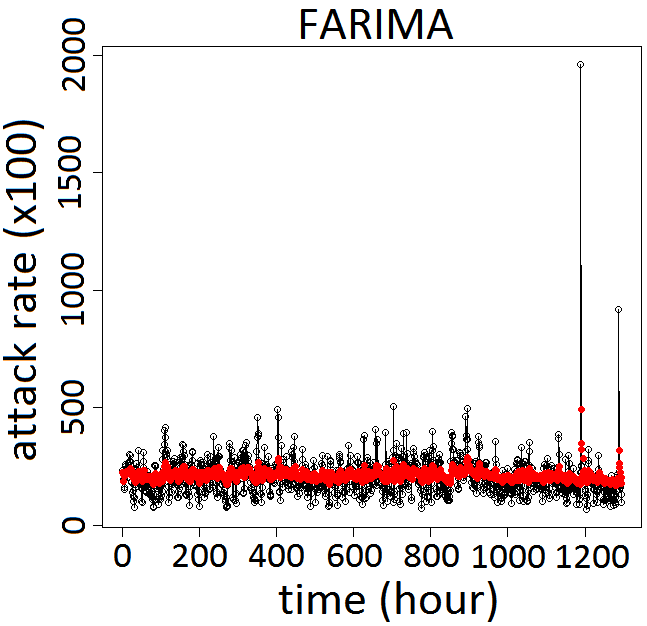}}
\subfigure[Period IV]{\includegraphics[width=.192\textwidth]{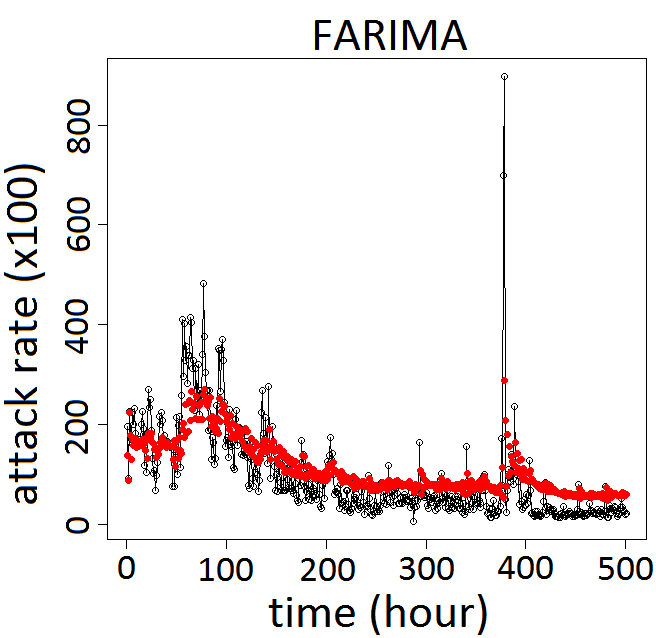}}
\subfigure[Period V]{\includegraphics[width=.192\textwidth]{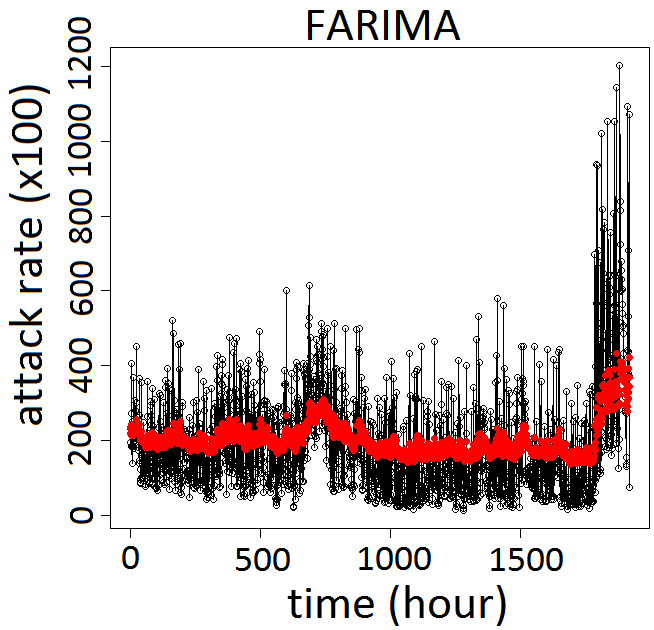}}
\caption{TST-based model fitting where black-colored circles represent the observed attack rates and
red-colored dots represent the fitted values. FARIMA+GARCH fits Periods I-III better than FARIMA (especially for the extreme attack rates), but cannot fit Periods IV-V well
(although FARIMA+GARCH fits more accurately than FARIMA).
\label{fig:fiting5}}
\end{figure*}

Figure \ref{fig:fiting5} plots the fitting results.
We observe that FARIMA+GARCH can indeed fit the data, especially the extreme attack rates, better than FARIMA.
For Periods I-III, we observe that the inaccuracy of FARIMA+GARCH is mainly because the extreme values (i.e., spikes) are not fitted 100\%.
For Period IV, Figure \ref{fig:fitting-d2-5d} visually shows that FARIMA+GARCH fits the data well.
However, the visual effect is misleading (meaning that purely visual analysis is not always reliable)
because there is some significant fitting error (as shown in Table \ref{table:garch+garch2} below).
Again, this is caused by the incapability in fitting extreme attack rates.
For Period V, FARIMA+GARCH clearly cannot fit the data well, possibly because there exist some complex statistical patterns that change rapidly.
We are not aware of any statistical tools that can cope with such time series, but expect to develop some advanced tools for this purpose in the future.

Table \ref{table:garch+garch2} summarizes the fitting results.
For Periods I-III, we observe that FARIMA+GARCH has smaller AIC values as well as smaller PMAD values (all $<0.2$).
For Periods IV-V, FARIMA+GARCH still has smaller PMAD and AIC values, but the PMAD values are greater than $0.3$.
Therefore, FARIMA+GARCH can fit Periods I-III better than FARIMA, but not well enough for Periods IV-V
(although FARIMA+GARCH fits these two periods more accurately than FARIMA).

\begin{table}[!htbp]
{\small
\centering
\begin{tabular}{|c|r|r|r|r|r|}
\hline
Period     & \multicolumn{3}{|c|}{${\sf FARIMA+GARCH}$} & \multicolumn{2}{|c|}{{\sf FARIMA}}   \\  \cline{2-5}\hline
 &  model & PMAD &AIC & PMAD & AIC  \\
\hline
{\rm I} & $M_1'$   &{\bf 0.170} & $11.0$   & $0.192$   & $11.8  $   \\ \hline
{\rm II} &$M_1'$   &{\bf 0.185} & $11.4 $  & $0.195$   & $11.9 $    \\ \hline
{\rm III} & $M_3'$ &{\bf 0.196}  & $ 10.7$ & $0.239$  & $  11.3$   \\ \hline
{\rm IV}  &$M_3'$  &$0.363 $ & $ 9.6 $ & $0.439 $ & $10.7$   \\ \hline
{\rm V} & $M_4'$   &$0.441$  & $11.8 $ & $0.482 $ & $12.0$   \\ \hline
\end{tabular}
\caption{TST-based fitting of attack rates (per hour). Gray-box FARIMA+GARCH models can
consistently fit Periods I-III well, and always fit better than gray-box FARIMA models. \label{table:garch+garch2}}
}
\end{table}

\section{Prediction Analysis}
\label{sec:connection}

\subsection{Gray-Box vs. Black-Box Prediction Accuracy}

In order to show the power of gray-box prediction,
we compare (i) the prediction accuracies of the gray-box FARIMA+GARCH models mentioned above
(all of these models are considered, rather than the best fitting models only, because the best fitting models may not always offer the best predictions),
(ii) the prediction accuracies of gray-box FARIMA models that can accommodate the LRD phenomenon but not the extreme-value phenomenon,
(iii) the prediction accuracies of black-box Hidden Markov Models \cite{zu2009hidden} (which are reviewed in Section \ref{sec:hmm}),
and (iv) prediction accuracy of black-box Symbolic Dynamics (SD) models \cite{lin2003symbolic} (which are reviewed in Section \ref{sec:sd}).

\paragraph{FARIMA+GARCH vs. FARIMA}
Recall that Table \ref{table:garch+garch2} shows that the gray-box FARIMA+GARCH models
offer more accurate fittings than gray-box FARIMA models.
Now we investigate to what extent FARIMA+GARCH can predict more accurately than FARIMA.
Algorithm \ref{alg:garch-based-prediction} describes the FARIMA+GARCH prediction algorithm, where the best prediction model $M'_j\in \{M'_1,\ldots,M'_4\}$ is selected,
and the last 120 hours are used for $h$-hour ahead-of-time prediction (which is consistent with the FARIMA-based prediction in \cite{XuIEEETIFS13}).

{\small
\begin{algorithm}[!hbpt]
\caption{FARIMA+GARCH-based prediction of attack rates}
\label{alg:garch-based-prediction}
INPUT: attack-rate time series $\{X_1,\ldots,X_n\}$, FARIMA-GARCH model family $\{M'_1, M'_2, M'_3, M'_{4}\}$, $0<\ell<1$, $h$ (the number of hours ahead-of-time prediction)\\
OUTPUT: best prediction model $M'\in \{M'_1,\ldots,M'_4\}$

\begin{algorithmic}[1]
\FOR{$i = 1$  \TO  $4$}
\STATE{$m=\lfloor n\ell \rfloor $, $j=0$}
\WHILE{$m+h\leq n$}
\STATE{use $\{X_1,\ldots,X_m\}$ to fit a $M'_i$-type model}
\STATE{use $M'_i$ to predict attack rates $\{X'_{m+1},\ldots,X'_{m+h}\}$}
\STATE{$m=m+h$}
\ENDWHILE
\STATE{evaluate PMAD value of the predictions}
\ENDFOR
\RETURN{$M'\in\{M'_1,M'_2,M'_3,M'_4\}$ with the smallest PMAD values}
\end{algorithmic}
\end{algorithm}
}

Table \ref{table:t-pred-d1-d2} summarizes the prediction results of the best prediction models.
We observe that for Periods I-III, 1-hour ahead-of-time predictions (i.e., $h=1$) lead to the highest prediction accuracy (at least 86\%).
Recall that FARIMA-based 1-hour ahead-of-time predictions for Periods I-V have PMAD values 0.179 (error: 17.9\%), 0.217 (error: 21.7\%), 0.298 (error: 29.8\%),
0.548 (error: 54.8\%), and 0.517 (error: 51.7\%), respectively \cite{XuIEEETIFS13}.
This shows that gray-box FARIMA+GARCH models can lead to an extra prediction accuracy of 4.1\%, 9.6\%, 15.8\%, 20.9\%, and 13.9\%, respectively.
We note that predictions for Periods IV and V are substantially less accurate than predictions for Periods I-III,
because their prediction errors are 33.9\% and 37.8\% respectively.
We suspect this discrepancy is caused by the data corresponding to Periods IV and V having some complex statistical properties (other than LRD and extreme values),
which are left for future investigations.
To summarize the above discussion, we conclude that
FARIMA+GARCH-based 1-hour ahead-of-time predictions offer the best accuracy, and we will use FARIMA+GARCH for comparison with HMM-based and SD-based predictions.

\begin{table}[htbp!]
{\footnotesize
\centering
\begin{tabular}{|c|r|c|r|r|r|r|}
\hline
\multirow{2}{*}{Period} & \multirow{2}{*}{$\ell$} & \multirow{2}{*}{Selected Model} & \multicolumn{4}{|c|}{PMAD} \tabularnewline
\cline{4-7}
 &  & & $h$=1H & $h$=4H & $h$=7H & $h$=10H\\
\hline
I  & $0.90$& $M_3'$ & {\bf 0.138} &0.172  &0.255 &0.300\\ \hline
II & $0.70$ & $M_4'$ & {\bf 0.121} &0.343 &0.390 &0.386\\ \hline
III& $0.90$ & $M_3'$ &{\bf 0.140}  &0.276 & 0.316 &0.282\\\hline
IV & $0.80$ & $M_3'$ &0.339 &0.409 &0.535 &1.152\\\hline
V  & $0.95$ & $M_3'$ &0.378 & 0.388 &0.470 &0.288\\\hline
\end{tabular}
\caption{Errors of the gray-box FARIMA+GARCH models for predicting the attack rate $h$-hour ahead-of-time: Periods I-III
can be predicted accurately 1-hour ahead-of-time.
\label{table:t-pred-d1-d2}}
}
\end{table}

\paragraph{FARIMA+GARCH vs. HMM (Hidden Markov Model)}
Algorithm \ref{alg:hmm-prediction} describes the HMM-based prediction algorithm with $i$ hidden states for $h$-hour ahead-of-time predictions.
The best prediction model is selected among the choices of different numbers of hidden states
(i.e., the number of hidden states that leads to the smallest PMAD value is selected).
To be specific, we consider $i\in\{2,3,\ldots,10\}$ hidden states.

{\small
\begin{algorithm}[!hbtp]
\caption{HMM-based predictions of attack rates}
\label{alg:hmm-prediction}
INPUT: Extreme attack rates $\{X_1,\ldots,X_n\}$; number of hidden states $i\in \{2, 3, \ldots, 10\}$;
 $0<\ell<1$, $h$ (\# of hours ahead-of-time prediction)\\
OUTPUT: best prediction model with $i$ hidden states

\begin{algorithmic}[1]
\FOR{$i = 2$  \TO  $10$}
\STATE{$m=\lfloor n\ell \rfloor $}
\WHILE{$m+h\leq n$}
\STATE{use $\{X_1,\ldots,X_m\}$ to fit a model of $i$ hidden state}
\STATE{use the fitted model to predict $\{X'_{m+1},\ldots,X'_{m+h}\}$}
\STATE{$m=m+h$}
\ENDWHILE
\STATE{evaluate PMAD value of the predictions}
\ENDFOR
\RETURN{$i\in\{2,3,\ldots,10\}$ with the smallest PMAD values}
\end{algorithmic}
\end{algorithm}
}

Table \ref{tab:HMM-one} compares HMM-based predictions of the last 120 hours of attack rates with that of FARIMA+GARCH-based predictions,
where $h=1,4,7,10$.
In most scenarios, FARIMA+GARCH-based predictions are substantially more accurate than HMM-based predictions.
For Periods I-III and $h=1$, FARIMA+GARCH-based predictions are respectively 0.8\%, 11.0\%, and 11.4\% more accurate than HMM-based predictions.
For Periods IV-V and $h=1$, FARIMA+GARCH-based predictions are respectively 21.2\% and 2.8\% more accurate than HMM-based predictions.
Despite that FARIMA+GARCH cannot predict Periods IV-V well, this model inadequacy does not invalidate
the gray-box prediction methodology. Instead, this suggests that we need to use even more sophisticated
models to accommodate the complex properties/phenomena exhibited by the Periods IV-V data.

\begin{table}[!htbp]
\centering
{\small
\begin{tabular}{|c|c|c|c|c|}
\hline
\multirow{2}{*}{Period} &  \multicolumn{4}{|c|}{PMAD} \tabularnewline
\cline{2-5}
 &   $h$=1H & $h$=4H & $h$=7H & $h$=10H\\
\hline
I & 0.146  & 0.193  & 0.189 & 0.191 \\
\hline
II & 0.231 & 0.342 & 0.371  & 0.349 \\
\hline
III & 0.254  & 0.379  & 0.401  & 0.427 \\
\hline
IV & 0.551  & 0.872  & 1.159  & 1.456 \\
\hline
V & 0.406  & 0.531  & 0.577  & 0.607 \\
\hline
\end{tabular}
\caption{Errors of
HMM-based $h$-hour ahead-of-time predictions: For 1-hour ahead-of-time prediction (which is the case we recommend for practical use),
HMM predictions are substantially less accurate than FARIMA+GARCH predictions.
\label{tab:HMM-one}
}}
\end{table}

\paragraph{FARIMA+GARCH vs. SD (Symbolic Dynamics) model}
The first step in using SD models is to
transform attack rates into symbols. For this purpose, we partition the attack rate range
into five intervals in terms of quantiles, namely $[0, 80\%\text{ quantile}]$, $(80\%\text{ quantile},85\%\text{ quantile}]$, $(85\%\text{ quantile}, 90\%\text{ quantile}]$,
$(90\%\text{ quantile}, 95\%\text{ quantile}]$, and $(95\%\text{ quantile}, 100\%\text{ quantile}]$.
These are represented as symbols 1, 2, 3, 4 and 5, respectively.
We consider the attack rates that are no greater than the 80\% quantile as a single interval
because we are analyzing the extreme attack rates.
For comparison purposes, the time resolution of the resulting symbolic time series is also an hour.
In order to predict the time series of symbols, we consider two models:
one is the aforementioned HMM, and the other is
the Conditional Random Field (CRF) model that selects symbols based on the maximum conditional probability \cite{lafferty2001d}.
Because SD-based predictions are also symbols, we need to map the predictions back into numerical attack rates.
Since each symbol corresponds to an interval, we consider three mappings:
(i) $min$-based mapping, in which case a symbol is mapped to the minimal value in the corresponding interval,
(ii) $mean$-based mapping, in which case a symbol is mapped to the mean value in the corresponding interval, and
(iii) $max$-based mapping, in which case a symbol is mapped to the maximum value in the corresponding interval.

{\small
\begin{algorithm}[!hbpt]
\caption{SD-based predictions of attack rates}
\label{alg:sym-prediction}
INPUT: Symbolic time series $\{Z_1,\ldots,Z_n\}$, number of hidden states $i\in\{2, 3, \ldots, 10\}$,
 $0<\ell<1$, $h$ ($h$-hour ahead-of-time prediction)\\
OUTPUT: best prediction models

\begin{algorithmic}[1]
\FOR{$i = 2$  \TO  $10$}
\STATE{$m=\lfloor n\ell \rfloor $}
\WHILE{$m+h\leq n$}
\STATE{use $\{Z_1,\ldots,Z_m\}$ to fit a HMM model of $i$ hidden states}
\STATE{use $\{Z_1,\ldots,Z_m\}$ to fit a CRF model of $i$ hidden states}
\STATE{use the fitted HMM model to predict symbols $\{Z'_{m+1},\ldots,Z'_{m+h}\}$}
\STATE{use the fitted CRF model to predict symbols $\{Z'_{m+1},\ldots,Z'_{m+h}\}$}
\STATE{$m=m+h$}
\ENDWHILE
\STATE{evaluate PMAD value of the predictions}
\ENDFOR
\RETURN{the best HMM model (i.e., the one with the smallest PMAD value among the HMM models), and the best CRF model (i.e., the one with the smallest PMAD values among the CRF models)}
\end{algorithmic}
\end{algorithm}
}

Algorithm \ref{alg:sym-prediction} is the SD-based prediction algorithm,
where $h=1$ (i.e., 1-hour ahead-of-time prediction) and we also use the last 120 hours of each period for prediction as in the preceding prediction algorithms
(i.e., the $\ell$ is calculated correspondingly).

Table \ref{tab:sym3} describes the PMAD-values of SD-based 1-hour ahead-of-time predictions.
We observe that $mean$-based mapping from symbols to numerical values offers relatively more accurate
predictions than $min$- and $max$-based mappings.
We also observe that for $mean$-based mapping, HMM and CRF perform almost the same.
However, neither HMM nor CRF provides accurate prediction in the SD framework
(recall that HMM alone can achieve a certain degree of accuracy, although not as accurately as FARIMA+GARCH).
Therefore, SD is not appropriate for predicting attack rates.
The fundamental reason CRF cannot predict well in this setting is that the data in question is univariate.
Despite the inaccuracy of SD, we keep these results for the sake of completeness.

\begin{table}[!htbp]
{\small
\center
\begin{tabular}{|c|c|c|c|c|c|c|}
\hline
\multirow{2}{*}{Period} &  \multicolumn{6}{|c|}{PMAD} \tabularnewline
\cline{2-7}
& $min$- & $mean$- & $max$- & $min$- & $mean$- & $max$- \tabularnewline
\hline
& \multicolumn{3}{c|}{ HMM (best model) } & \multicolumn{3}{c|}{ CRF (best model) }\tabularnewline
\hline
I & 0.754 & 0.457 & 0.613 & 0.754 & 0.457 & 0.613\tabularnewline
\hline
II & 0.754 & 0.336 & 0.375 & 0.591 & 0.364 & 0.404\tabularnewline
\hline
III & 0.635 & 0.309 & 0.367 & 0.629 & 0.316 & 0.369\tabularnewline
\hline
IV & 0.775 & 0.553 & 0.808 & 0.775 & 0.553 & 0.808 \tabularnewline
\hline
V & 0.920 & 0.528 & 0.754 & 0.920 & 0.528 & 0.754 \tabularnewline
\hline
\end{tabular}
\caption{\label{tab:sym3}Errors of SD-based 1-hour ahead-of-time predictions: The predictions are inaccurate and described here for completeness.
}
}
\end{table}

\paragraph{Summary}

Based on the above discussion, we conclude that gray-box FARIMA+GARCH models, which can accommodate both the LRD phenomenon and the extreme-value phenomenon, can predict
attack rates 1-hour ahead-of-time at an accuracy that is substantially greater than that of
gray-box FARIMA models, which can accommodate the LRD phenomenon but not the extreme-value phenomenon.
Moreover, gray-box FARIMA+GARCH predictions are substantially more accurate than HMM predictions, and are far more accurate than SD predictions.
Therefore, we will compare FARIMA+GARCH predictions to EVT-based predictions of expected magnitude of extreme attack rates.

\subsection{EVT-based prediction of extreme attack rates}

We use EVT-based methods to predict the return level,
namely the expected {\em magnitude} of extreme attack rates within a future period of time
(but not necessarily the expected {\em maximum} attack rate).
We use Algorithm \ref{alg:evt-based-prediction} to predict return levels for the return period of $h$ hours (i.e., $h$-hour ahead-of-time predictions).
Since EVT-based predictions only deal with extreme values, we cannot use PMAD to measure prediction errors.
Instead, we use the {\em binomial test}  \cite{MCNEIL2000} to measure the prediction accuracy.
A $p$-value greater than .05 indicates that a prediction is accurate.
For comparison purposes, we set $h=24$ and use the last 120 hours in each period for prediction
(i.e., $\ell$ is chosen such that $m=\lfloor n\ell \rfloor=n-120$).

{\small
\begin{algorithm}[!hbtp]
\caption{EVT-based prediction of return levels}
\label{alg:evt-based-prediction}
INPUT: extreme attack rates $\{X_1,\ldots,X_n\}$ with respect to threshold quantiles described in Tables \ref{table:basic-stat-stationary}-\ref{tbl:non-stationary-d1d2},
EVT model family $\{M_1, M_2, M_3, M_{4}\}$ described in Section \ref{sec:evt-analysis},
$0<\ell<1$, $h$ (\# of hours as return period)\\
OUTPUT: prediction model $M_j\in\{M_1,M_2,M_3,M_4\}$
\begin{algorithmic}[1]
\STATE{$m=\lfloor n\ell \rfloor$}
\FOR{$i = 1$  \TO  $4$}
\WHILE{$m+h\leq n$}
\STATE{Using $\{X_1,\ldots,X_m\}$ to fit the $M_i$-type model}
\STATE{Use $M_i$ to predict the return level between the $(m+1)$th and the $(m+h)$th hours}
\STATE{$m=m+h$}
\ENDWHILE
\STATE{Evaluate prediction accuracy using the binomial test}
\ENDFOR
\RETURN{$M_j \in\{M_1,M_2,M_3,M_4\}$ with the highest $p$-value (or simpler model with the same $p$-value)}
\end{algorithmic}
\end{algorithm}
}

\subsection{Making use of EVT- and TST-based predictions}

Table \ref{table:leveld1} reports EVT-based predictions of return levels as well as the corresponding $p$-values of the binomial test,
the observed maximum attack rates, and the TST-based predictions of maximum attack rates
 (more precisely, FARIMA+GARCH predictions) as well as the corresponding PMAD values.
We make the following observations.

\begin{table}[hbtp!]
\centering
{\footnotesize
\begin{tabular}{|c|r|r|r|r|r|r|}
\hline
\multirow{2}{*}{Per.} & \multirow{2}{*}{H1-24} & \multirow{2}{*}{H25-48} & \multirow{2}{*}{H49-72} & \multirow{2}{*}{H73-96} & \multirow{2}{*}{H97-120} & $p$ or \tabularnewline
 &  &  &  &   &  & PMAD\\ \hline
\multirow{3}{*}{I}   & 76056 & 76824 & 77088 & 78333 & 82275 &  {\bf 0.07} \\ \cline{2-7}
                     & 53197 & 60203 & 57370 & 62868 & 82576 & \\ \cline{2-7}
                     & 50656 & 53744 & 52427 & 58183 & 71719 &{\bf 0.13}\\\hline
\multirow{3}{*}{II}  &  60910 & 60668  & 63572 & 62750 & 60752 &  {\bf 0.17}\\ \cline{2-7}
                     &  83157 & 101937 & 45186 & 38218 & 60274 & \\ \cline{2-7}
                     &  72457 & 83073  & 43942 & 37267 & 51853 & {\bf 0.18}\\\hline
\multirow{3}{*}{III} &  33263 & 32916 & 32836 & 32733 & 32654 &  {\bf 0.07}\\\cline{2-7}
                     &  32993 & 30194 & 21476 & 92379 & 29722 & \\ \cline{2-7}
                     &  30869 & 28505 & 21382 & 77747 & 27921 & {\bf 0.21}\\\hline
\multirow{3}{*}{IV}  &  29747 & 29622 & 28622 & 30048 & 30514 & {0.01}\\ \cline{2-7}
                     &  23224 & 11671 & 6634 & 6263 & 8225 & \\ \cline{2-7}
                     &  {\em 20329} & {\em 9443}  & {\em 6674} & {\em 5341} & {\em 5605} & $0.40$\\\hline
\multirow{3}{*}{V}  &  40129 & 41265  &  42360  &  43526  &  45996  & 0.00   \\ \cline{2-7}
                    & 101971 & 105171 & 114462  & 120221  & 109216  &        \\ \cline{2-7}
                    &  40341 & 43581  &  50053  &  54644  &  55007  & 0.41 \\\hline
 \end{tabular}
\caption{Comparison between EVT- and TST-based predictions, where ``H$a$-$b$" means the predictions correspond to time interval between the $a$th and the $b$th hours
(among the last 120 hours of each period).
For each period, there are three rows.
The first row represents EVT-based predictions of return levels (i.e., expected magnitude of extreme attack rates) and the corresponding $p$-value of the predictions,
where the prediction model
is selected by Algorithm \ref{alg:evt-based-prediction}.
The second row describes the observed (i.e., actual) maximum attack rates.
The third row describes the maximum attack rates derived from the TST-based predictions of attack rates made by
the FARIMA+GARCH model
selected by Algorithm \ref{alg:garch-based-prediction}
with $h=1$ and the corresponding PMAD value.
\label{table:leveld1}}
}
\end{table}

First, EVT-based best prediction models (corresponding to the first row of each period in Table \ref{table:leveld1}) are respectively simpler than
the best EVT-based fitting models
(described in Tables \ref{table:basic-stat-stationary}-\ref{tbl:non-stationary-d1d2}).
This means that the best fitting models are not necessarily the best prediction models.
On the other hand, TST-based best prediction models (corresponding to the third row in each period in Table \ref{table:leveld1}) are respectively the
same as the TST-based best fitting models (described in Table \ref{table:t-pred-d1-d2}).

Second, we observe a consistency between the predictions of these two approaches.
Specifically, EVT-based predictions of return levels (i.e., expected magnitude of extreme attack rates)
are accurate for Periods I-III because their $p$-values are greater than 0.05,
while TST-based predictions of the maximum attack rates are also accurate for Periods I-III because their PMAD values are smaller than or equal to 0.22.
On the other hand, EVT-based predictions of return levels are not accurate for Periods IV-V because their $p$-values are smaller than 0.02,
and TST-based predictions of the maximum attack rates are not accurate for Periods IV-V because their PMAD values are greater than or equal to 0.40.
However, there is a significant difference between Periods IV and V.
For Period IV, we observe that TST-based predictions of the maximum attack rates
are actually accurate with respect to the observed
maximum attack rates. This suggests that although TST-based predictions are not accurate overall, their predictions of the maximum attack rates can be accurate.
This is useful because the predicted maximum attack rates can be the most important factor for
the defender's decision-making for  resource allocation.
Unfortunately, TST-based predictions of the maximum attack rates are not accurate for Period V.
This may be caused by some complex time series patterns, such as cyclical trends or seasonal trends (i.e., some repeated patterns).
We leave the settlement of this issue to future investigations.

\begin{figure*}[!hbtp]
\centering
\subfigure[Period I]{\includegraphics[width=.192\textwidth]{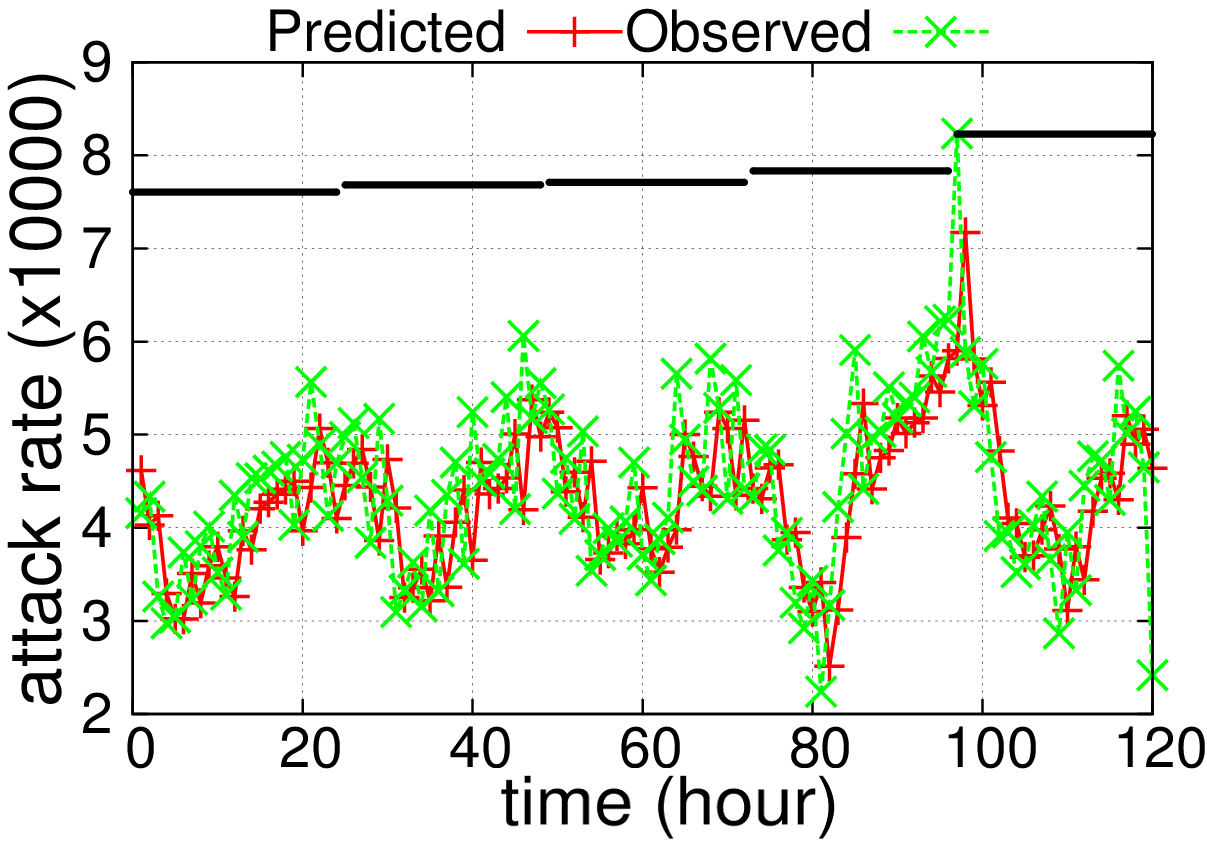}}
\subfigure[Period II]{\includegraphics[width=.192\textwidth]{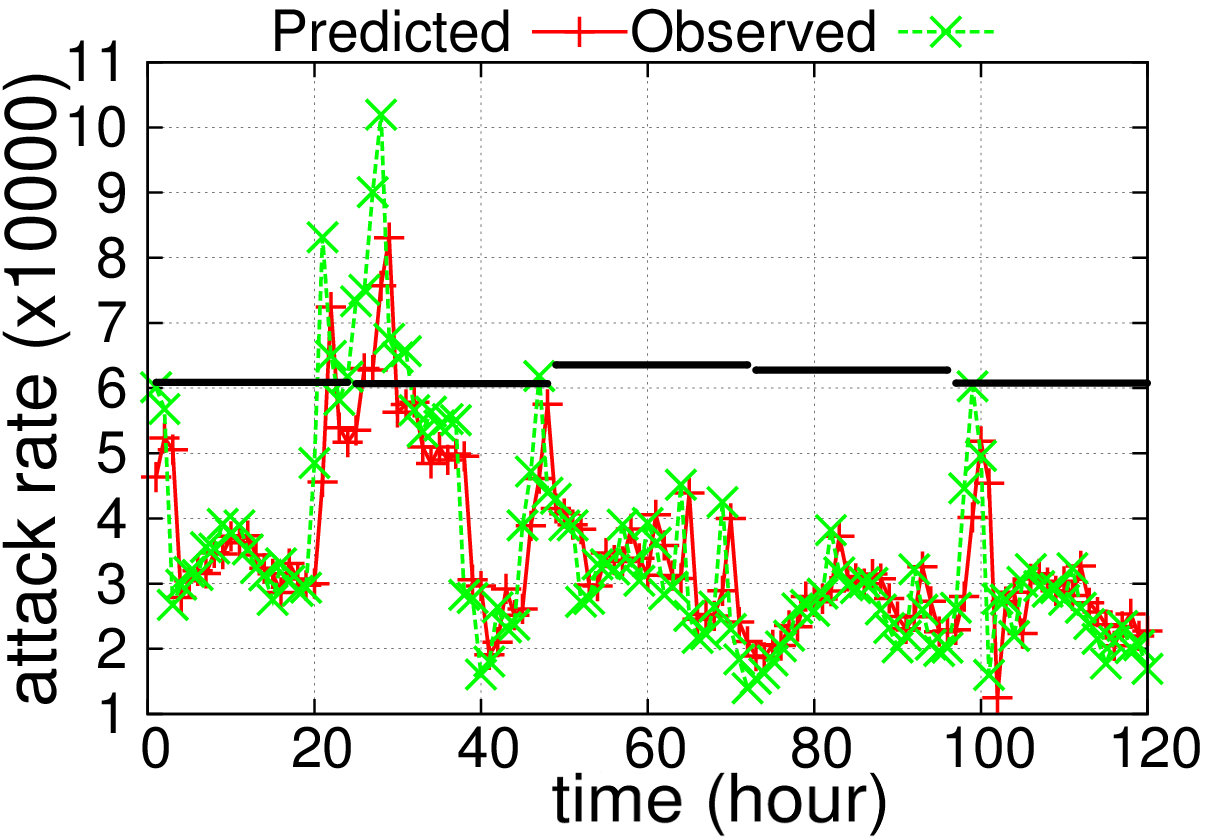}}
\subfigure[Period III]{\includegraphics[width=.192\textwidth]{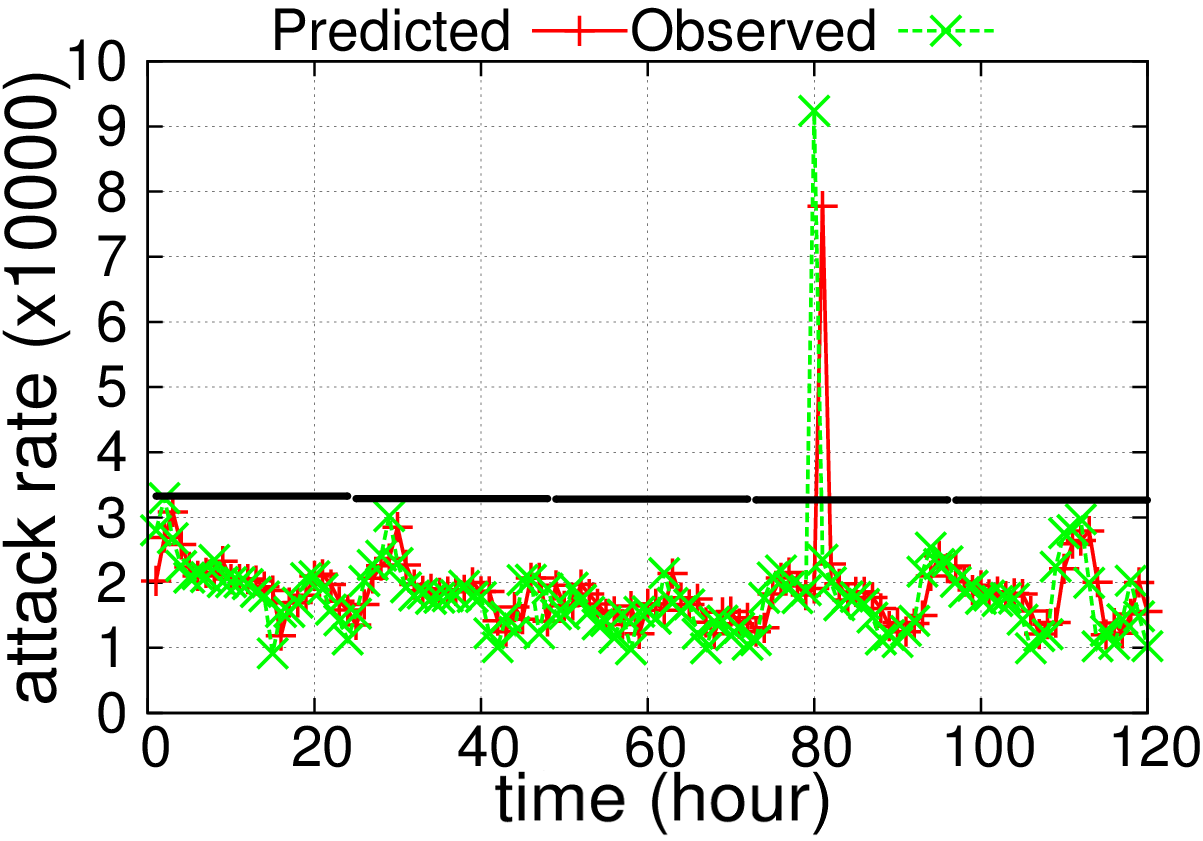}}
\subfigure[Period IV]{\includegraphics[width=.192\textwidth]{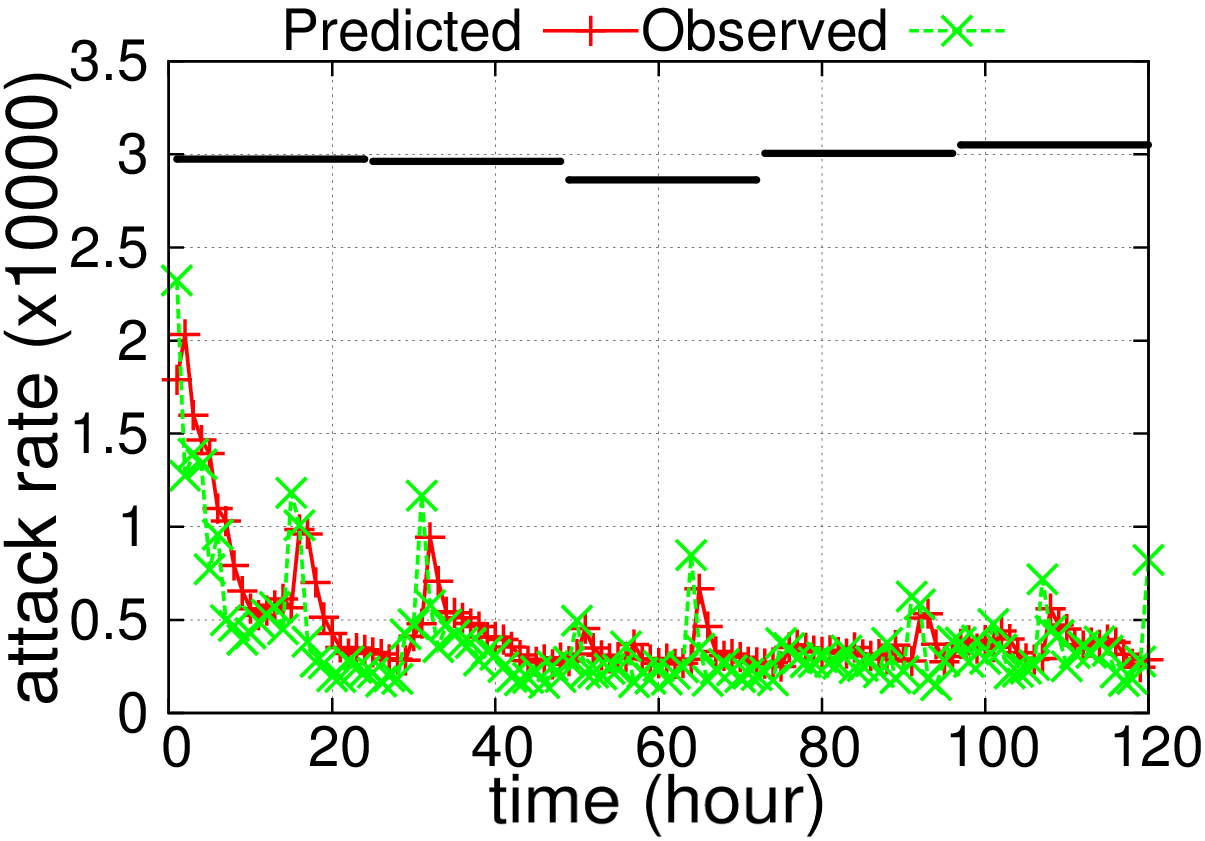}\label{fig:prediction120-d1}}
\subfigure[Period V]{\includegraphics[width=.192\textwidth]{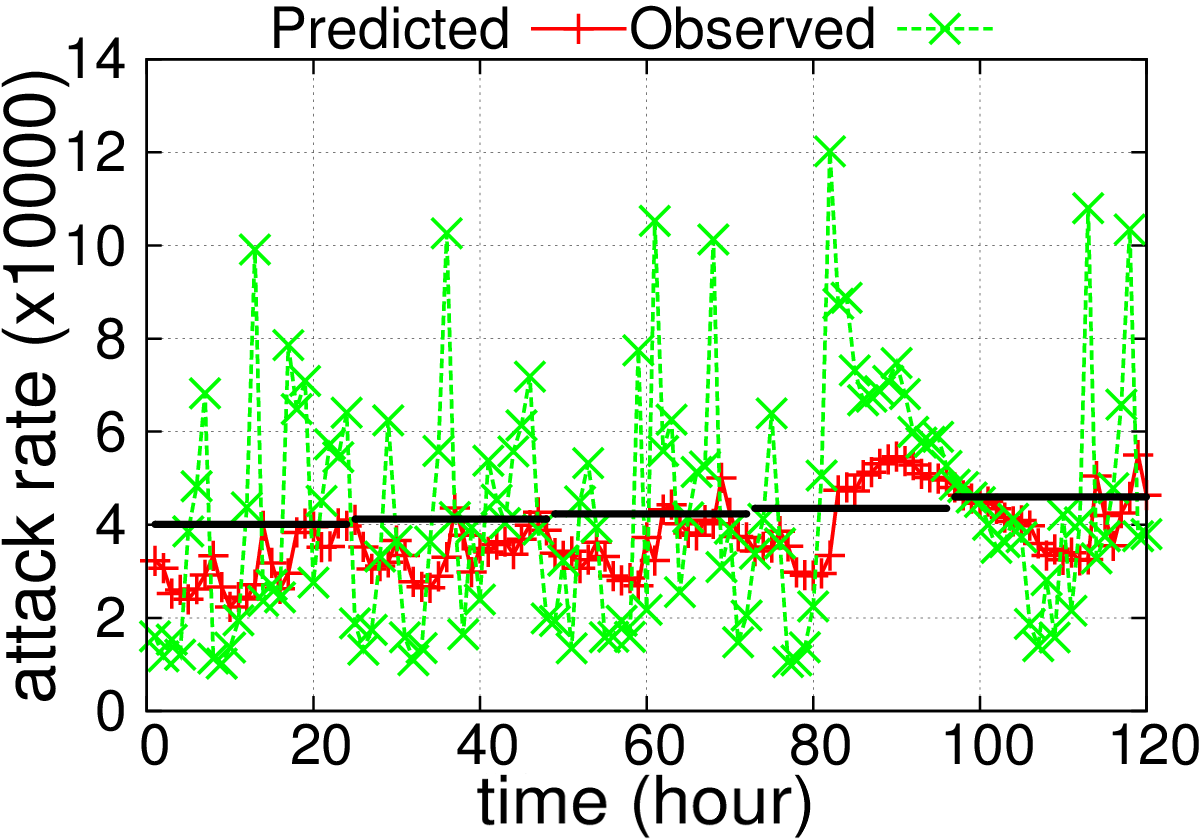}}
\caption{Comparing EVT-based predictions of return levels (i.e., expected magnitudes of extreme attack rates), observed attack rates during
the last 120 hours in each period, and TST-based predictions of attack rates.
EVT-based predictions of return levels are produced by Algorithm \ref{alg:evt-based-prediction} and summarized in Table \ref{table:leveld1},
and plotted as horizontal lines during the respective intervals of 24 hours.
TST-based predictions are produced by Algorithm \ref{alg:garch-based-prediction}.
For Periods I-III, EVT-based predictions of return levels are accurate,
and TST-based predictions of attack rates as well as the {\em maximum attack rates} are also accurate.
For Period IV, EVT-based predictions of extreme attack rates are about one order of magnitude above the observed attack rates,
but TST-based predictions of maximum attack rates are accurate.
For Period V, neither EVT nor TST can predict accurately.
\label{fig:prediction120}}
\end{figure*}

Third, for Periods I-III where both EVT and TST provide overall accurate predictions, we observe that EVT-based predictions of return levels, which are
given to the defender 24 hours ahead-of-time, can be used as an evidence for resource allocation.
Moreover, EVT-based resource allocations could be dynamically adjusted by taking into account TST-based 1-hour ahead-of-time predictions.
Specifically, Figure \ref{fig:prediction120}
suggests the following: when TST-based prediction of the maximum attack rate (which is obtained only 1 hour ahead-of-time) is above EVT-based prediction of the
return level (which is obtained 24 hours ahead-of-time),
the defender can dynamically allocate further resources for the anticipated attack rate predicted by TST-based methods (e.g., the highest spike in
Periods II-III as shown in Figure \ref{fig:prediction120}). Clearly, this strategy is conservative because it (on average) overprovisions resources to cope with the worst-case
scenario (i.e., matching the highest attack rate).
Nevertheless, this strategy gives the defender more early-warning time.
An alternative strategy is to use TST-based prediction of the maximum attack rate as the initial evidence for allocating defense resources,
then take into consideration EVT-based long-term predictions (e.g., some weighted average). This strategy might prevent resource overprovisioning,
but may not provision sufficient resources for coping with the highest attack rate (e.g., the highest spike in Period III as shown in Figure \ref{fig:prediction120}).
This strategy requires the defender to be more agile (than in the preceding strategy).

\section{Limitations and Future Research Directions}
\label{sec:limitations}

The present study has the following limitations, which nevertheless suggest directions for future research.
First, the analysis results are limited by the data, but are sufficient for justifying the value of the gray-box prediction methodology and the newly introduced
family of FARIMA+GARCH models.
Indeed, gray-box FARIMA+GARCH predictions are substantially more accurate than gray-box FARIMA predictions,
which in turn are much more accurate than black-box ARMA predictions \cite{XuIEEETIFS13}.
The analysis methodology can be equally applied to analyze other data sets of a similar kind.

Second, we observe that EVT-based predictions of return levels are often above the actual maximum attack rates,
but TST-based predictions of the maximum attack rates are often below the actual maximum attack rates.
This connection between EVT and TST is just a good starting point.
For example, it may be possible to use some weighted average of EVT-based predictions of return levels and TST-based predictions of maximum attack rates
as the predicted maximum attack rate.
These heuristics needs to be justified rigorously.
Moreover, there might be some deeper connections that can be exploited to formulate more powerful prediction techniques.
Finally, there may be some fundamental trade-off between the early-warning time we can give to the defender
and the prediction accuracy we can expect.
These connections have not be investigated by the theoretical statistics community,
and our engineering-driven demand might give theoretical statisticians enough motivation to explore this exciting topic.

Third, Periods IV and V cannot be predicated accurately,
possibly because there exist some properties other than LRD and extreme values.
Further studies are needed in order to determine, for example, if there are some {\em seasonal} or {\em cyclical} trends,
and/or if the extreme values are generated by some self-exciting process \cite{mcneil2010quantitative}.

\section{Conclusions}
\label{sec:conclusion}

We have presented a novel methodology for analyzing the extreme-value phenomenon exhibited by cyber attack data.
The analysis methodology utilizes a novel integration of EVT and TST, and aims to predict attack rates more accurately by accommodating the extreme-value phenomenon.
We have presented gray-box FARIMA+GARCH models,
which can accommodate both the LRD phenomenon and the extreme-value phenomenon that are exhibited by the data,
and can predict attack rates 1-hour ahead-of-time at an accuracy that can be deemed practical.
We believe that this study will inspire an exciting research sub-field, including the adequate treatment of the open problems mentioned above.

\smallskip

\noindent{\bf Acknowledgement.} We thank the reviewers for their constructive comments that helped us  improve the paper,
including the comparison of FARIMA+GARCH-based predictions to HMM-based and SD-based predictions.
We thank the associate editor, Professor Wanlei Zhou, for his constructive comments.
We thank Marcus Pendleton and Paul Parker for proofreading the paper.

This study was IRB-approved.
This work was supported in part by ARO Grant \#W911NF-13-1-0141.
Any opinions, findings, and conclusions or recommendations expressed in this material are those of
the author(s) and do not necessarily reflect the views of the funding agencies.

\begin{wrapfigure}{l}{0.15\textwidth}
\centering
\includegraphics[height=0.12\textwidth,width=0.12\textwidth]{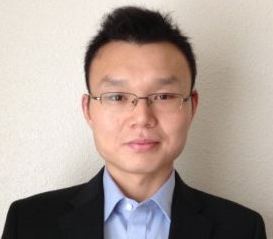}
\end{wrapfigure}
Zhenxin Zhan recently received his PhD in cyber security from the Department of Computer Science, University of Texas at San Antonio.
He received M.S. degree in Computer Science from the Huazhong University of Science and Technology, China, in 2008.
His primary research interests are in cyber attack analysis and detection.

\begin{wrapfigure}{l}{0.15\textwidth}
\centering
\includegraphics[height=0.12\textwidth,width=0.12\textwidth]{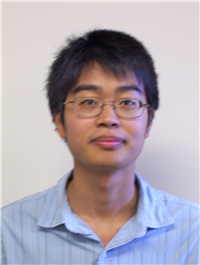}
\end{wrapfigure}
Maochao Xu is an assistant professor of Mathematics at the Illinois State University.
He received his PH.D. in Statistics from Portland State University in 2010.
His research interests include Applied Statistics, Extreme value theory,  Cyber security, and Risk analysis in actuary and insurance.
He currently serves as an associate editor for Communications in Statistics.

\begin{wrapfigure}{l}{0.15\textwidth}
\centering
\includegraphics[height=0.12\textwidth,width=0.12\textwidth]{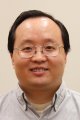}
\end{wrapfigure}
Shouhuai Xu is a full professor in the Department of
Computer Science, University of Texas at San Antonio.
His research interests include cybersecurity modeling \& analysis (especially, Cybersecurity Dynamics) and applied cryptography.
He earned his PhD in Computer Science from Fudan University, China.
He is an associate editor for IEEE TDSC and IEEE T-IFS.
More information about his research can be found at
\url{www.cs.utsa.edu/~shxu}.

\end{document}